\documentclass[amssymb,amsmath,prd,twocolumn,superscriptaddress,nofootinbib]{revtex4-1}

\usepackage{graphicx}
\usepackage{bm}
\usepackage{amssymb,amsmath}
\usepackage{mathrsfs}
\usepackage{latexsym}
\usepackage{color}
\usepackage[normalem]{ulem}
\usepackage{dcolumn}
\usepackage[colorlinks=true,citecolor=blue,urlcolor=blue]{hyperref}
\usepackage[usenames,dvipsnames]{xcolor}
\usepackage{xspace}
\usepackage{graphicx}

\allowdisplaybreaks

\newcommand{\CCA}{\affiliation{Center for Computational Astrophysics, Flatiron Institute, 162 5th Ave, New York, NY 10010, USA}}

\newcommand{\MSFC}{\affiliation{NASA Marshall Space Flight Center, Huntsville, AL 35812, USA}}

\newcommand{\XGI}{\affiliation{eXtreme Gravity Institute, Department of Physics, Montana State University, Bozeman, Montana 59717, USA}}

\newcommand{\CIT}{\affiliation{Department of Physics, California Institute of Technology, Pasadena, California 91125, USA}}
\newcommand{\CITLab}{\affiliation{LIGO Laboratory, California Institute of Technology, Pasadena, CA 91125, USA}}
\newcommand{\US}{\affiliation{Mathematical Sciences and STAG Research Centre, University of Southampton, SO17 1BJ, Southampton, UK}}

\definecolor{kcmagenta}{rgb}{0.54, 0.17, 0.88}
\definecolor{chorange}{rgb}{0.851, 0.372, 0.007}
\definecolor{tlteal}{rgb}{0,.55,.55}
\definecolor{jcpink}{rgb}{1.0, 0.0, 0.5}
\definecolor{mmgreen}{rgb}{0.0, 0.8, 0.6}
\definecolor{bbsalmon}{rgb}{1.0, 0.47, 0.42}

\newcommand{\BayesLine}{{\tt BayesLine}\xspace}
\newcommand{\BayesWave}{{\tt BayesWave}\xspace}

\newcommand{\GlitchBuster}{{\tt GlitchBuster}\xspace}
\newcommand{\QuickCBC}{{\tt QuickCBC}\xspace}

\newcommand{\chieff}{\chi_{\textrm{eff}}}

\begin{document}

\title{Modeling compact binary signals and instrumental glitches in gravitational wave data}

\author{Katerina Chatziioannou} \CIT \CITLab \CCA
\author{Neil J. Cornish} \XGI
\author{Marcella Wijngaarden}  \CCA\US
\author{Tyson B. Littenberg}\MSFC

\date{\today}

\begin{abstract}
Transient non-Gaussian noise in gravitational wave detectors, commonly referred to as glitches, pose challenges for detection and inference of the astrophysical properties of detected signals when the two 
are coincident in time.
Current analyses aim toward modeling and subtracting the glitches from the data using a flexible, morphology-independent model in terms of sine-Gaussian wavelets 
before the signal source properties are inferred using templates for the compact binary signal. 
We present a new analysis of gravitational wave data that contain both a signal and glitches by simultaneously modeling the compact binary signal in terms of templates 
and the instrumental glitches using sine-Gaussian
wavelets. The model for the glitches is generic and can thus be applied to a wide range of glitch morphologies without any special tuning. The simultaneous modeling of the astrophysical signal
with templates allows us to efficiently separate the signal from the glitches, as we demonstrate using simulated signals injected around real O2 glitches in the two LIGO detectors.
We show that our new proposed analysis can separate overlapping glitches and signals, estimate the compact binary parameters, and provide ready-to-use glitch-subtracted data for downstream 
inference analyses.
\end{abstract}

\maketitle

\section{Introduction}
\label{sec:intro}

During the first half of their third observing run (O3a), the advanced ground based gravitational wave (GW) detectors LIGO~\cite{TheLIGOScientific:2014jea} and Virgo~\cite{TheVirgo:2014hva} 
observed an astrophysical transient signal about every 5 days of data~\cite{Abbott:2020niy}.
The large detection rate increases the chance of observing an event while one of the detectors experiences transient non Gaussian noise, also known as instrumental glitches.
Indeed, this scenario has come to pass for one event from the second observing run (O2)~\cite{TheLIGOScientific:2017qsa} and 8 events from the first
half of the third observing run~\cite{Abbott:2020niy}.

Such coincidences are expected to become even more frequent in the coming years.
Planned improvements in the detectors' sensitivity will be directly reflected by an even larger rate 
of astrophysical discoveries~\cite{Aasi:2013wya}. Moreover, O3a was characterized by an increase in the rate of glitch occurrence in the two LIGO detectors, 
a trend that might persist during the fourth observing run (O4) as the decreased average detector noise might help reveal weaker sources of transient noise. 
For example the rate of glitches in the LIGO Livingston detector increased from 0.2 per minute in O2 to 0.8 per minute in O3a~\cite{Abbott:2020niy}.

The presence of a non Gaussian noise feature in the data, a glitch, poses challenges for nearly all inference analyses. GW inference is based on a model
for the detector noise, expressed through the likelihood function. In the absence of glitches, detector noise is colored and Gaussian to a very good approximation~\cite{Chatziioannou:2019}, 
with a spectrum 
that is described through the noise power spectral density (PSD). The above considerations give rise to a Gaussian likelihood function whose variance is the noise PSD, a choice that is
almost ubiquitous~\cite{Veitch:2014wba,LIGOScientific:2019hgc}. 
Different choices for estimating the PSD or treating its uncertainty can result in different functional forms for the likelihood, but they are all based on the
assumption of colored Gaussian noise~\cite{Rover:2008yp,Talbot:2020auc}.

Since instrumental glitches violate the basic assumptions of GW inference, they need to be effectively mitigated before the data are analyzed. One option is to remove the offending data all
together~\cite{Usman:2015kfa,TheLIGOScientific:2017qsa,Sachdev:2019vvd,Zackay:2019kkv}, which can be done quickly, allowing for low latency estimation of source parameters that enable followup
observations~\cite{TheLIGOScientific:2017qsa}. The downside of this approach is that part of the astrophysical signal is lost making it prohibitive for binary black hole (BBH) signals whose duration
is comparable to the glitch duration. In order to avoid signal, and thus information loss, another option is to model the glitch and regress it from the data, leaving behind not only the
astrophysical signal but also the Gaussian noise. This approach is the topic of the current study\footnote{An independent effort to mitigate the effect of broadband and/or nonstationary 
detector noise is based on information from auxiliary sensors~\cite{2012CQGra..29u5008D,Tiwari:2015ofa,Meadors:2013lja,Driggers:2018gii,Davis:2018yrz,Vajente:2019ycy,Ormiston:2020ele}. This approach does not remove entire data segments either and thus is not expected to lead to loss of information.}.

The wide variety of glitch morphologies, and even variations within a certain glitch type, make constructing exact models for glitches challenging~\cite{Coughlin:2019ref}. A more flexible approach is based 
on \BayesWave~\cite{Cornish:2014kda,Cornish:2020dwh} which models various components of the GW data in a morphology-independent way. Non-Gaussian features in the data are
modeled in terms of sums of sine-Gaussian wavelets whose number and parameters are marginalized over with a suite of Markov Chain Monte Carlo (MCMC) 
and Reversible Jump MCMC (RJMCMC)~\cite{10.1093/biomet/82.4.711}
samplers. Coherent features (i.e. features that appear in all detectors in a manner consistent with an astrophysical signal originating from a specific sky location) are modeled by a single
sum of wavelets that is projected onto the detector network; these features are interpreted as having an astrophysical origin. Incoherent features are instead modeled by independent
sums of wavelets in each GW detector and are interpreted as instrumental glitches.
The PSD of the Gaussian noise is also modeled in terms of splines and Lorentzians using an algorithm sometimes known as \BayesLine~\cite{Littenberg:2014oda,Chatziioannou:2019}. \BayesWave and \BayesLine are fully integrated and we will  refer to the combined analysis with the name \BayesWave in this paper.

Modeling instrumental glitches with \BayesWave and subtracting them from the data in order to make ready-to-use data for downstream inference has been a 
standard step of LIGO/Virgo analyses since O2~\cite{TheLIGOScientific:2017qsa,Abbott:2020niy}. The GW signal from the first binary neutron star (BNS) coalescence detection, GW170817,
overlapped with a glitch in the LIGO Livingston detector approximately $1.1$s before coalescence~\cite{TheLIGOScientific:2017qsa}. The glitch was modeled with 
\BayesWave's \emph{glitch model} in terms of a sum of wavelets and removed from the data, a procedure documented and released in~\cite{GW170817Data}. Despite the glitch overlapping
with the actual astrophysical signal, the subtraction process was robust against inadvertently removing the signal together with the glitch. The reason is that the specific glitch was short in duration
(less than a second) and extended in frequency, unlike the signal that lasted for about $2$ minutes in the detector sensitive band. As such, the sine-Gaussian wavelets that would fit the glitch 
and the signal are distinct in terms of their time-frequency features; the wavelets that model the glitch are short and hence do not model the long-lasting BNS signal. This procedure
was further shown to not introduce biases in the astrophysical parameter inference of the underlying signal by analyzing simulated signals injected on instances of the same glitch type
in LIGO Livingston data~\cite{Pankow:2018qpo}.

Motivated by the success of this first attempt at glitch mitigation and in preparation for the increased detection rate of O3, \BayesWave was extended to be able to simultaneously model both 
the signal and the glitch~\cite{Cornish:2020dwh}. Both signals and glitches are modeled with a sum of sine-Gaussian wavelets, the only difference being that the signal is coherent across
the detectors in the network, while the glitch is not. The analysis effectively uses data from all detectors available to determine which part of the non Gaussian data are coherent 
(and would thus correspond to an astrophysical signal), and which part is incoherent (and would thus correspond to an instrumental glitch).
The combined signal+glitch analysis was applied to one O3a detection~\cite{Abbott:2020niy}, enabling glitch mitigation even for data that contained short-duration BBH signals.

The signal+glitch analysis models compact binary coalescence (CBC) signals in terms of wavelets, and is thus agnostic to the signal morphology. However, accurate models
exist for CBCs in terms of solutions to the Einstein field equations that are routinely used both for detection and parameter estimation. In this paper we take another step toward
efficient separation of CBCs and glitches by constructing an analysis that simultaneously models the CBC signal in terms of CBC templates and the glitch in terms
of sine-Gaussian wavelets. Similar to the initial glitch-only analysis and the subsequent signal+glitch analysis, we also model and marginalize over the detector noise PSD.
We test our analysis using public O2 data that contain common glitch types and simulated CBC signals.
We demonstrate that we can efficiently separate the glitch from the CBC, estimate the CBC parameters, and provide ready-to-use glitch-subtracted data for downstream 
inference analyses.

The rest of the paper is organized as follows. In Sec.~\ref{sec:method} we describe the updates to the standard \BayesWave algorithm in terms of the 
CBC analysis.
In Sec.~\ref{sec:inj} we apply our analysis to simulated signal overlapping with known detector glitches from O2 data. In Sec.~\ref{sec:events} we analyze
a selection of detected signals, namely GW170817 and GW150914.
Finally, in Sec.~\ref{sec:conclusions} we conclude and point to future work.

\section{General Algorithm Description}
\label{sec:method}

The combined \BayesWave algorithm is presented in detail in~\cite{Cornish:2020dwh} and here we describe only the features relevant to our study. \BayesWave simultaneously models
signals, glitches, and Gaussian noise in GW data by means of different models. 
The \emph{signal model} describes astrophysical signals through a sum of Morlet Gabor wavelets that are coherent across the detector network. The number
of wavelets and the parameters of each are marginalized over, as are the extrinsic parameters that determine how the signal is projected in each detector.
The \emph{glitch model} describes instrumental glitches with an incoherent sum of Morlet Gabor wavelets whose number and parameters are again marginalized over. 
Glitch power in each detector is described by an independent sum of such wavelets.
The \emph{noise model} describes the Gaussian noise PSD with a broadband spline model and sharp Lorentzians. As above, the number of spline points and Lorentzians
as well as their parameters are marginalized over.

In order to sample the multidimensional posterior density of all models, \BayesWave uses a blocked Gibbs sampler that takes turns between sampling each model with completely independent
 MCMC or RJMCMC samplers.
This includes (i) an RJMCMC that samples the signal and glitch wavelet parameters, 
(ii) an MCMC that samples the signal extrinsic parameters, and (iii) an RJMCMC that samples the splines and 
Lorentzians for the noise PSD. Each sampler in turn updates its parameters for a predetermined number of iterations, typically ${\cal{O}}(10^2)$, while all
other parameters are kept fixed. For example, the extrinsic sampler updates the extrinsic signal parameters while the wavelet parameters and noise PSD are kept
constant. Once the predetermined number of updates has been reached, 
the extrinsic sampler returns its current parameters and the noise sampler begins updating the
noise model while keeping the wavelet and extrinsic parameters fixed. This process of alternating sampling between different blocks of model parameters is repeated for ${\cal{O}}(10^4)$ iterations. 

The construction of the algorithm in terms of a blocked Gibbs sampler makes adding further models and samplers straightforward. In the current version described in~\cite{Cornish:2020dwh}, 
the astrophysical signal is modeled with
coherent sine-Gaussian wavelets that allow us to describe signals with a large level of flexibility. We extend \BayesWave's blocked Gibbs sampler by adding one more element, namely a model of
the signal in terms of quasicircular CBC waveforms. In fashion with the existing implementation, the MCMC that samples the posterior distribution for the CBC parameters is completely
independent from the remaining code samplers. The result is a flexible algorithm that can be used with any combination of CBC, signal\footnote{We retain the original model names in \BayesWave, hence 
the \emph{signal model} refers to the wavelet signal model, while the \emph{CBC model} refers to the model in terms of CBC templates. Both models target astrophysical signals. Since we do not use the \emph{signal model}
in the remainder of the paper, we trust that this will not lead to confusion.}, glitch, and noise models for the detector data.

The CBC model is integrated with {\tt LALSimulation}~\cite{lalsuite} and can operate with any nonprecessing model available there\footnote{Both the sampling and the jump proposals for the CBC parameters are constructed to expect the signal amplitude and phase from the waveform generator. There is therefore no fundamental limitation to non-precessing signals and
we plan to extend our analysis to include the effect of spin-precession in the future.}. The eleven parameters of a spin-aligned quasicircular CBC signal, namely the four intrinsic parameters (the two masses and spin magnitudes) and seven extrinsic parameters (the time of coalescence, the phase of coalescence,  two sky location angles, the polarization angle and the inclination angle the distance), are updated in overlapping blocks. Common to both blocks is the phase of coalescence since \BayesWave's extrinsic sampler updates the overall phase of the signal as described in~\cite{Cornish:2020dwh}.
 The CBC MCMC sampler updates the four intrinsic parameters, the time of coalescence, the phase of coalescence, and the distance. The existing extrinsic sampler in \BayesWave updates the two sky angles, the polarization angle, the inclination angle, and the phase of coalescence while holding all other parameters fixed. We use standard priors for all parameters: uniform over the detector-frame masses and spin magnitudes, uniform in time and phase, and uniform in luminosity volume. 

The CBC sampler is custom and not based on any existing samplers used in LIGO-Virgo parameter estimation. The CBC sampler is taken from the recently developed \QuickCBC~\cite{Cornish:2021wxy} analysis pipeline. A closely related sampler~\cite{Cornish:2020vtw} has been developed for analyzing data from the future Laser Interferometer Space Antenna. The CBC sampler is a replica exchange (parallel tempered) Markov Chain Monte Carlo (PTMCMC) algorithm that uses a mixture of proposal distributions. The default collection of proposals are: Gaussian jumps along eigenvectors of the Fisher information matrix, scaled by the reciprocal of the square root of the corresponding eigenvalue; differential evolution using a rolling history array at each temperature, updated every 10 iterations and holding 1000 past samples; and small, Gaussian jumps along each parameter direction. Each chain carries its own Fisher information matrix, which is updated periodically. The Fisher and differential evolution proposals are effective at exploring parameter correlations, while the small jumps prevent the chains from getting stuck in regions where the Fisher matrix becomes ill-conditioned.

The CBC sampler is not optimized for blindly finding signals, so it is best to initialize the sampler with a good starting solution for the source parameters such as the output from a CBC search pipeline, or the injected parameters for a simulated signal. Alternatively the sampler can be initialized using a custom built CBC search algorithm from the \QuickCBC~\cite{Cornish:2021wxy} analysis pipeline that has been incorporated into the \BayesWave preprocessing steps. The search is broken into two stages, a rapid network-coherent search with analytic maximization over extrinsic parameters, followed by a fast MCMC over the extrinsic parameters using a likelihood function that precomputes the waveform inner products~\cite{Cornish:2016pox}. This procedure returns the starting point for all $11$ CBC parameters.
More details about the initial search step and discussion of its robustness against instrumental glitches are presented in~\cite{Cornish:2021wxy}.

\section{Simulated Signals}
\label{sec:inj}

\begin{figure*}
\includegraphics[width=0.32\textwidth]{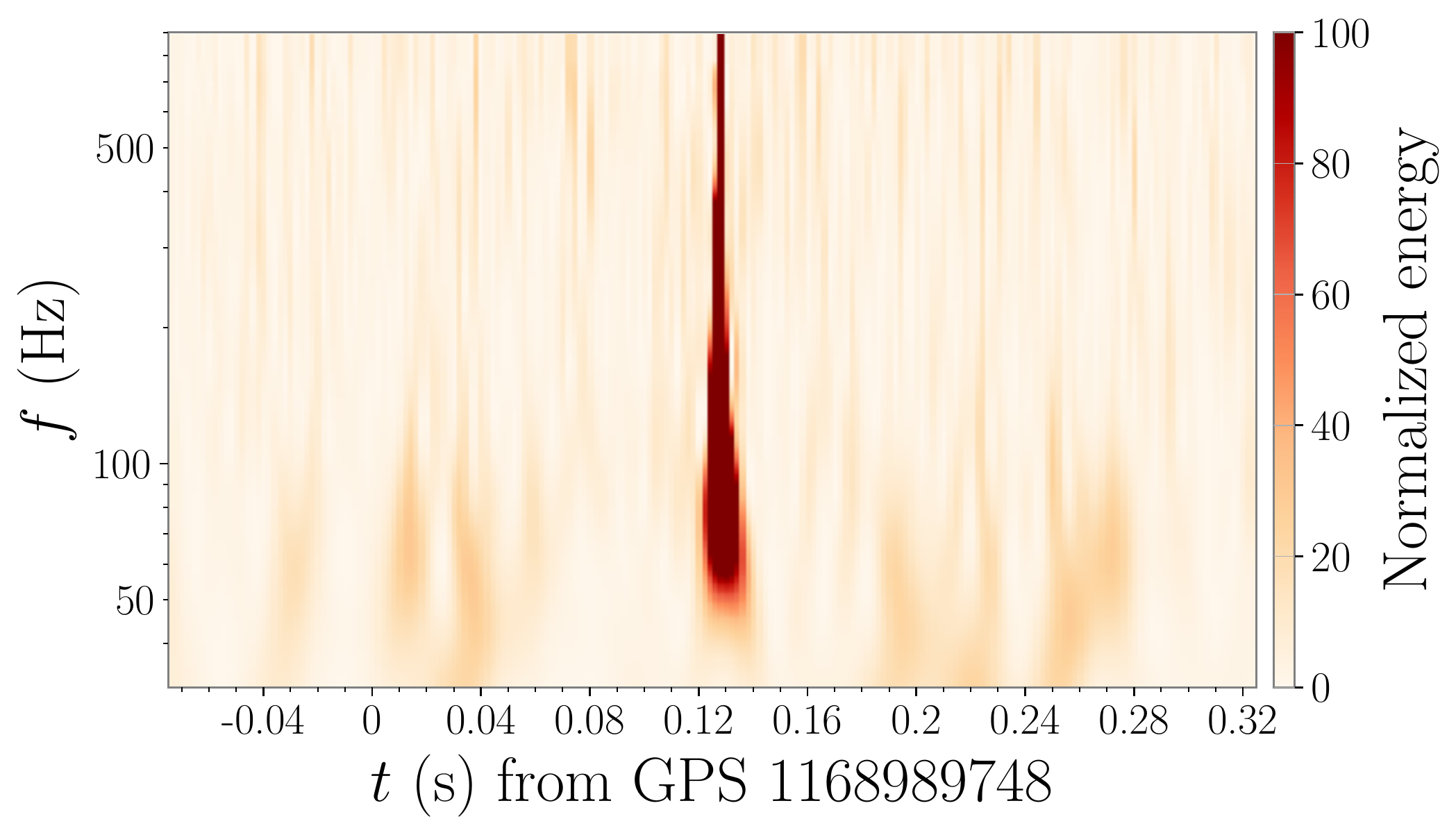}
\includegraphics[width=0.32\textwidth]{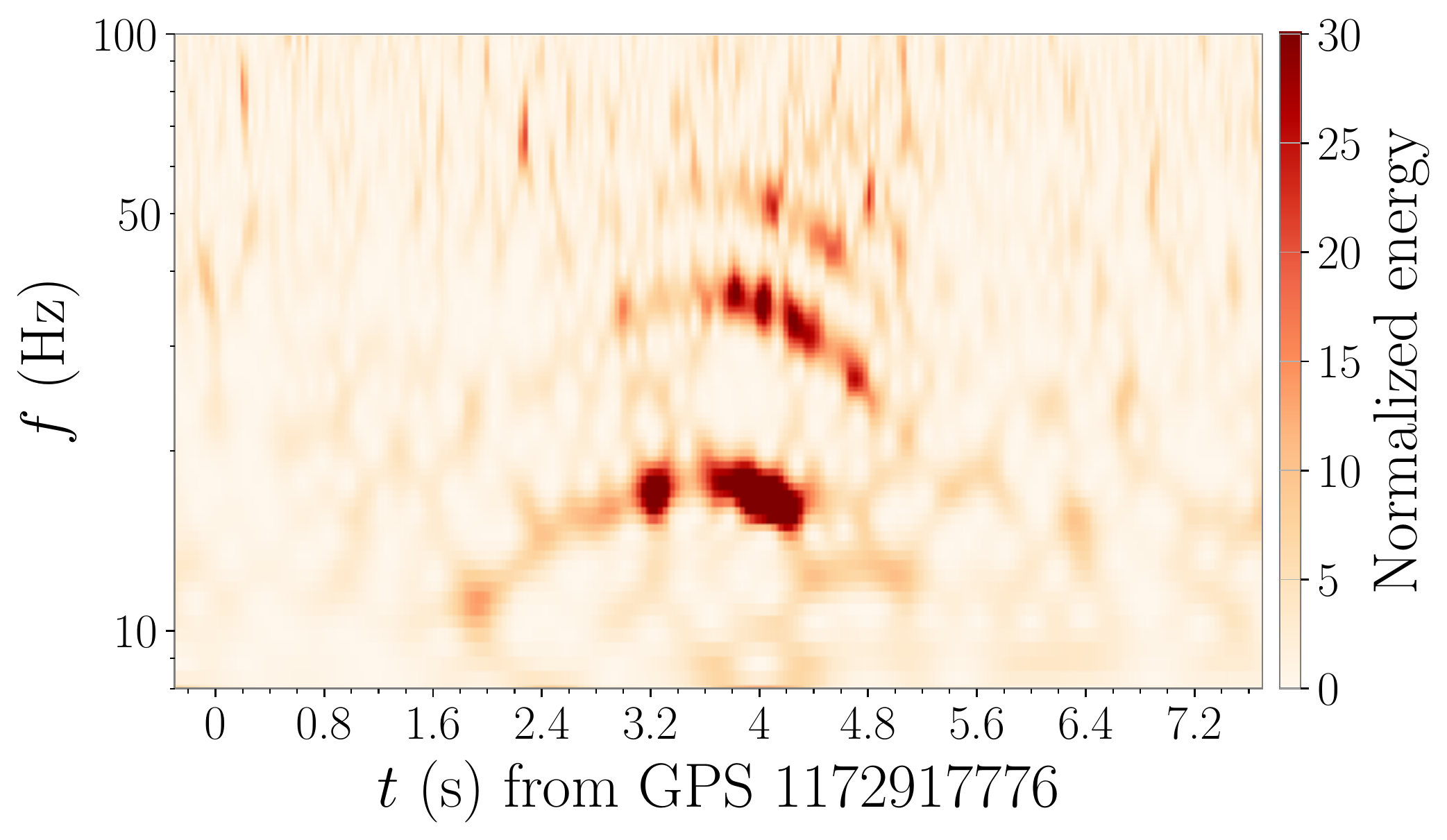}
\includegraphics[width=0.32\textwidth]{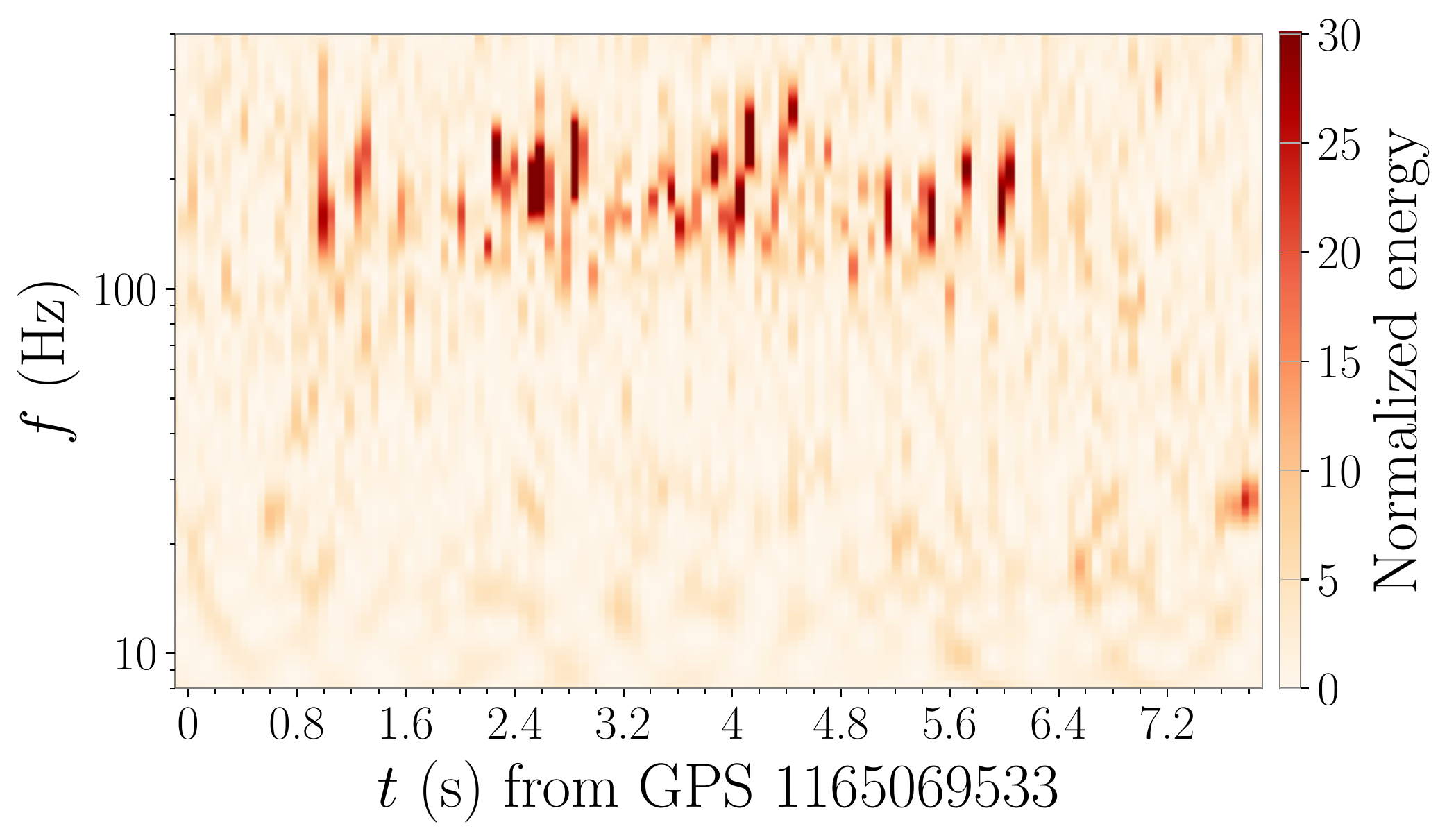}
\caption{Spectrograms for the three glitches of different types studied here: blip glitch (left), scattered light (middle), blue mountain (right). The three types of 
glitches are characterized by very
different time-frequency properties.}
\label{fig:SpecAll}
\end{figure*}

\begin{table*}
\centering
\begin{tabular}{c|c|c|c|c|c|c|c} 
Glitch & GPS time (s) &  Detector&\begin{tabular}[c]{@{}c@{}}Segment \\length (s) \end{tabular} &  \begin{tabular}[c]{@{}c@{}}Sampling \\ rate (Hz)\end{tabular}&  $f_{\textrm{low}}$(Hz) 
& $Q_{\textrm{max}}$ & CBC SNR \\ 
\hline
Blip & 1168989748  &Hanford & 4 &  2048 & 16 & 40 & 15\\
Scattered light  & 1172917779   & Livingston& 8 & 2048 & 8 & 160 & 15\\
Blue mountain & 1165069536 & Hanford &16  & 2048  & 16 & 40 & 15\\
\hline
\end{tabular}
\caption{
Settings for the runs of Sec.~\ref{sec:inj}. From left to right, columns correspond to the type of glitch, the GPS time, the affected detector, the segment length, the sampling rate, 
to low frequency cut off, the maximum quality factor of the glitch wavelets, and the SNR of the injected signals. 
}
\label{tab:settings}
\end{table*}

We test the efficacy of separating CBCs from glitches with our CBC+glitch model by selecting $3$ common glitch types from O2 data~\cite{GWOSC,Abbott:2019ebz}
that are known to have an adverse effect on searches for CBCs~\cite{Abbott:2020niy}.
 We then add simulated 
CBC signals consistent with a BBH with detector-frame masses of $36M_\odot$ and $29M_{\odot}$ and vanishing spin at different times with respect to the glitch. All simulated signals have a signal-to-noise (SNR) ratio of 15.
We use the {\tt IMRPhenomD}~\cite{Husa:2015iqa,Khan:2015jqa} waveform model both for simulation and recovery as implemented in 
{\tt LALSimulation}~\cite{lalsuite}.
We then analyze the data from the two LIGO detectors 
with our CBC+glitch+noise model, where the coherent signal is modeled by the CBC template, the glitch is modeled by incoherent 
wavelets, and the noise PSD is modeled with splines and Lorenzians. 
Spectrograms for the $3$ glitches are shown in Fig.~\ref{fig:SpecAll}: blip glitch (left), scattered light (middle), and blue mountain (right).
Further details and run settings for each type of glitch are shown in Table~\ref{tab:settings}.

\subsection{Glitch type 1: Blip}

\begin{figure*}
\includegraphics[width=\textwidth]{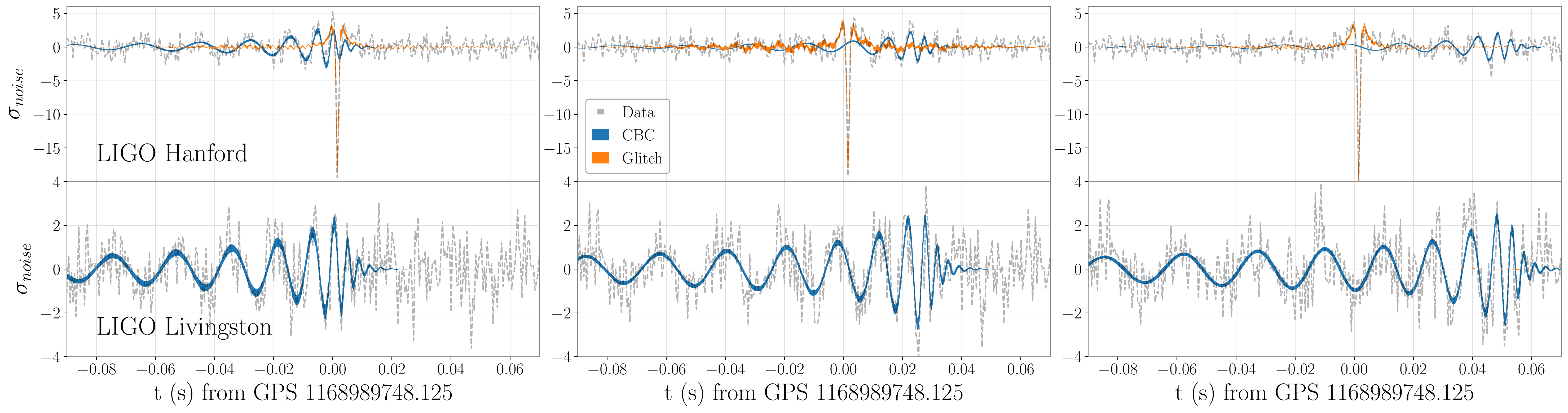}
\caption{Credible intervals for the glitch (orange) and the CBC (blue) signal reconstruction for data containing a blip glitch in LIGO Hanford and a simulated CBC signal at $3$ different 
times with respect to the glitch (left to right). Shaded regions correspond to 90\% credible intervals for the whitened reconstruction, 
while in grey dashed lines we plot the data whitened with a fair draw PSD from our
noise model posterior. The top row corresponds to LIGO Hanford and the bottom row corresponds to LIGO Livingston.}
\label{fig:BlipRec}
\end{figure*}

\begin{figure}
\includegraphics[width=0.49\textwidth]{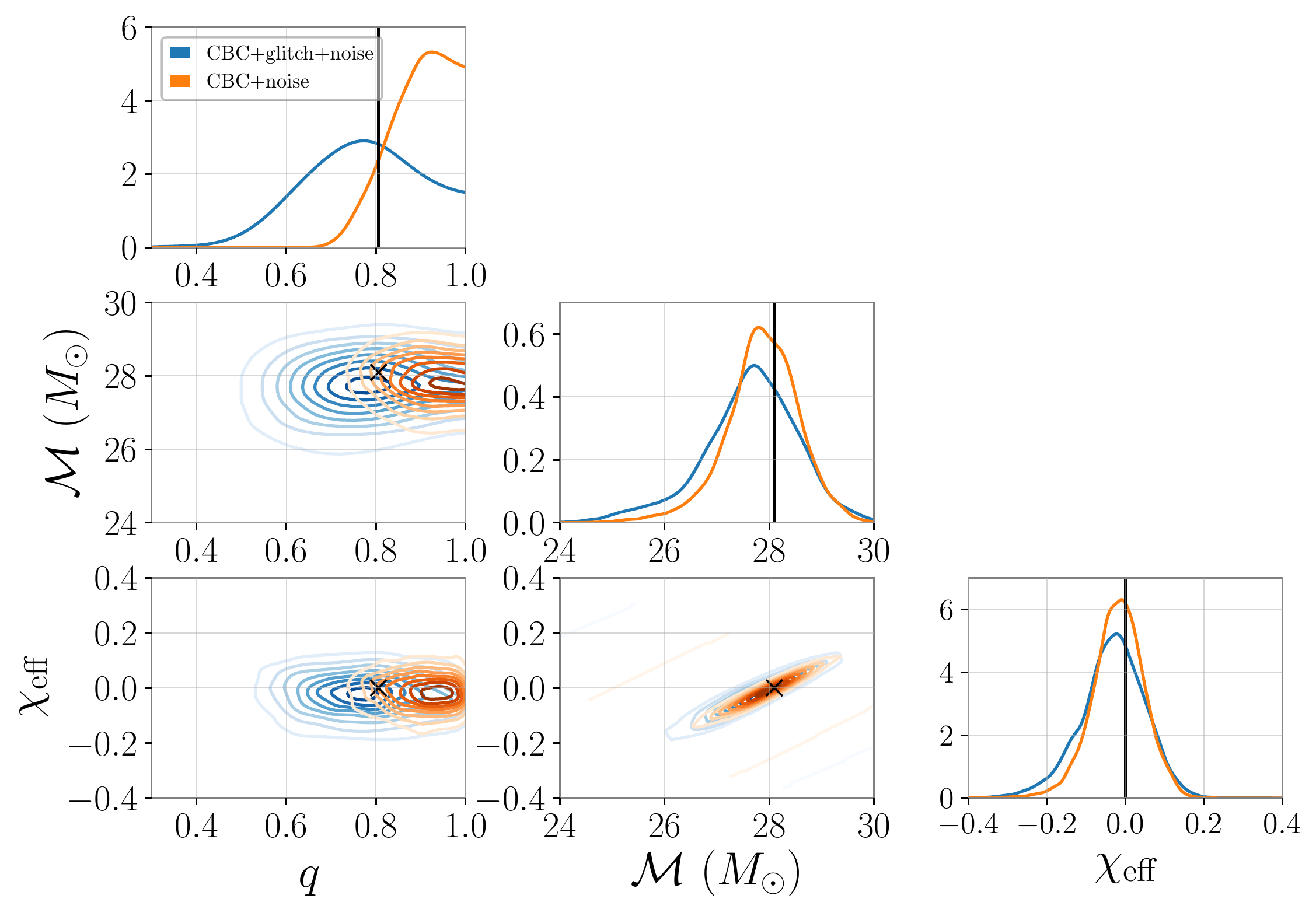}
\caption{One- and two-dimensional posterior distributions for selected source parameters of the simulated 
signal from the left panel of Fig.~\ref{fig:BlipRec} injected on top of a
LIGO Hanford blip glitch. We include the mass ratio $q$, the effective spin $\chieff$, and the detector frame chirp mass ${\cal{M}}$ posteriors, while black crosses or black vertical lines
denote the true parameters of the injection. Blue (orange) contours and lines correspond to the CBC+glitch+noise (CBC+noise) run.}
\label{fig:BlipParams}
\end{figure}

\begin{figure}
\includegraphics[width=0.49\textwidth]{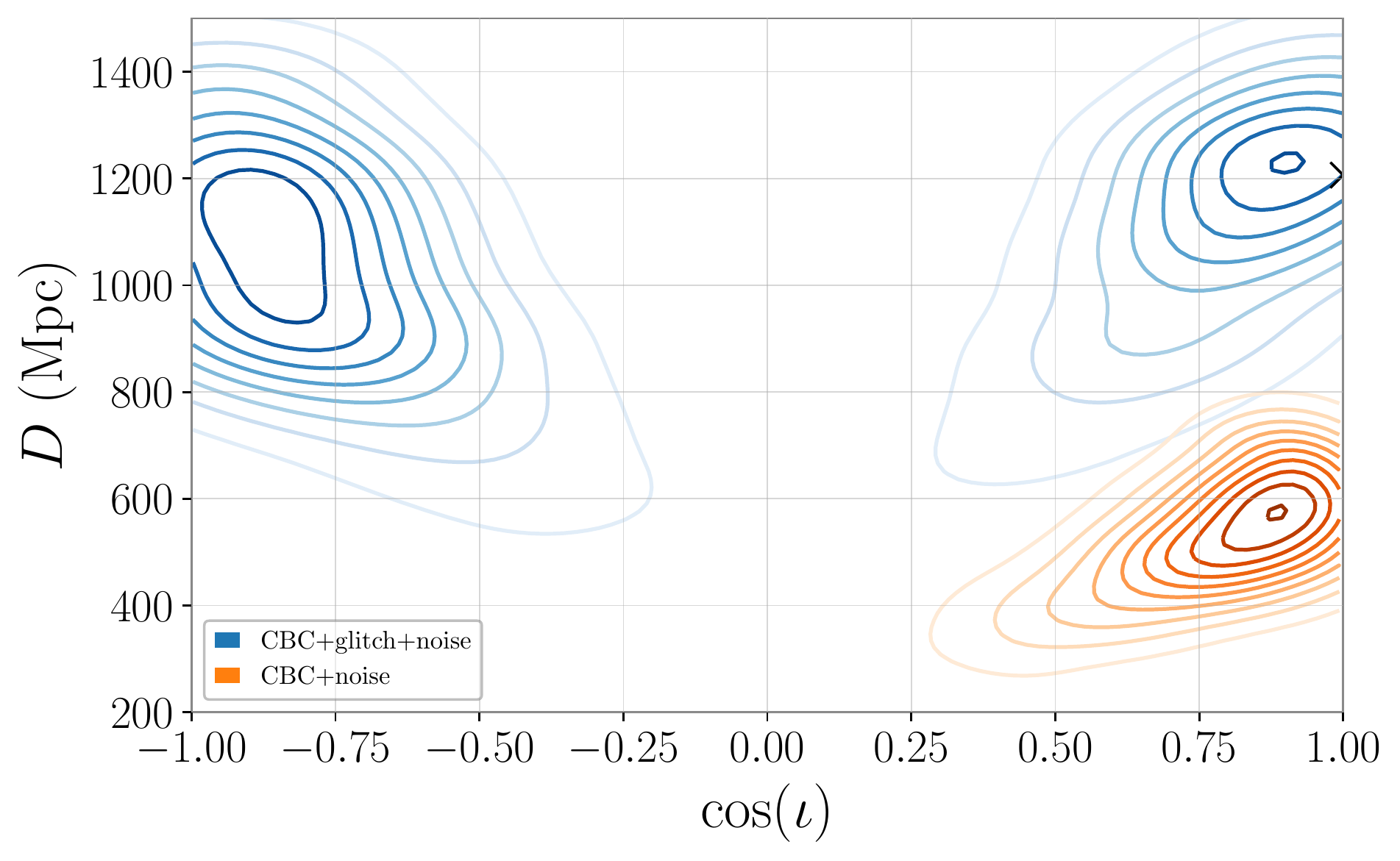}
\caption{Two-dimensional posterior distributions for the luminosity distance and the binary inclination of the simulated 
signal from the left panel of Fig.~\ref{fig:BlipRec} injected on top of a
LIGO Hanford blip glitch. A black cross at (1,1200Mpc) 
denotes the true parameters of the injection. Blue (orange) contours correspond to the CBC+glitch+noise (CBC+noise) run.}
\label{fig:BlipParams_Di}
\end{figure}

\begin{figure*}
\includegraphics[width=0.32\textwidth]{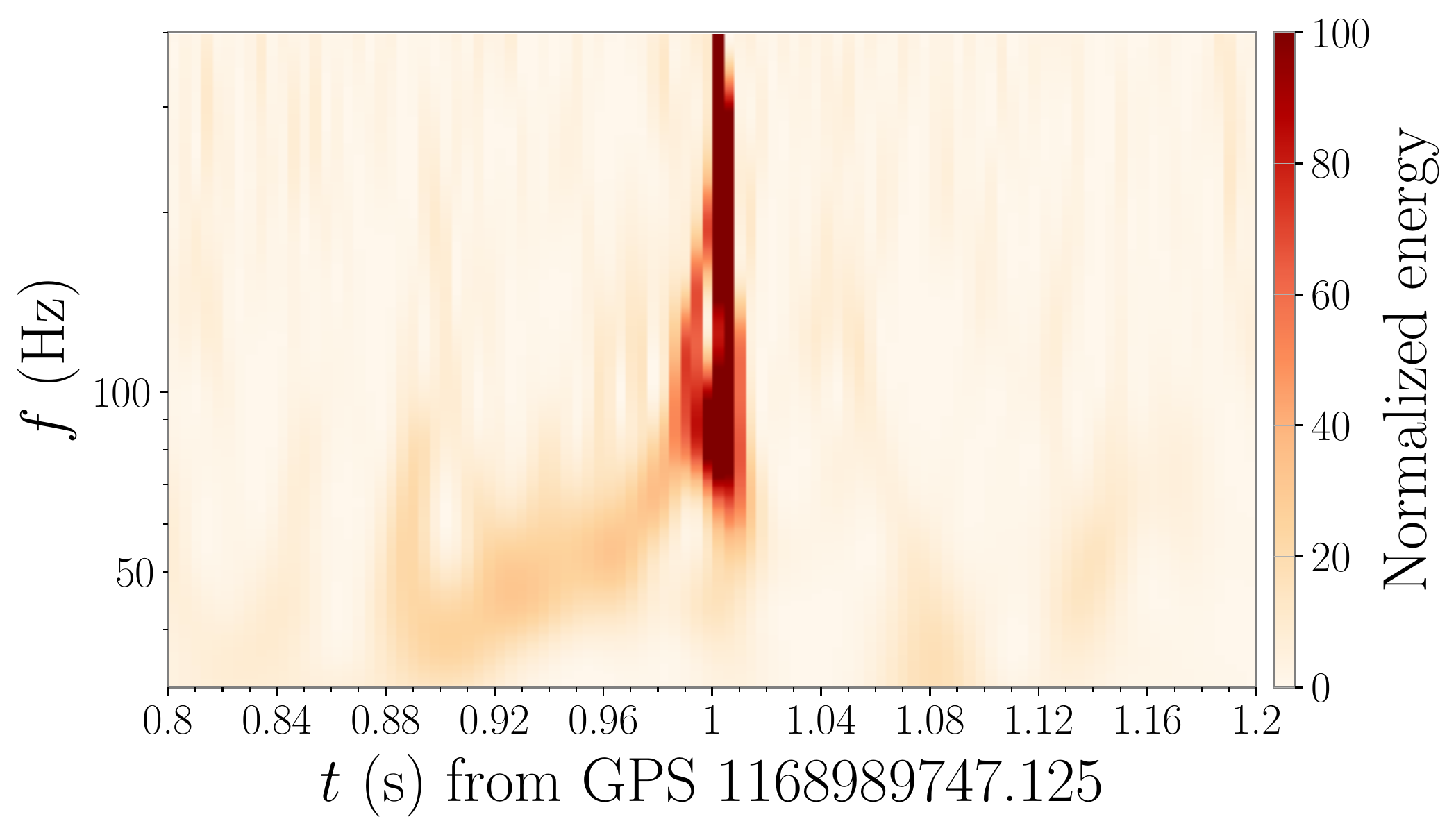}
\includegraphics[width=0.32\textwidth]{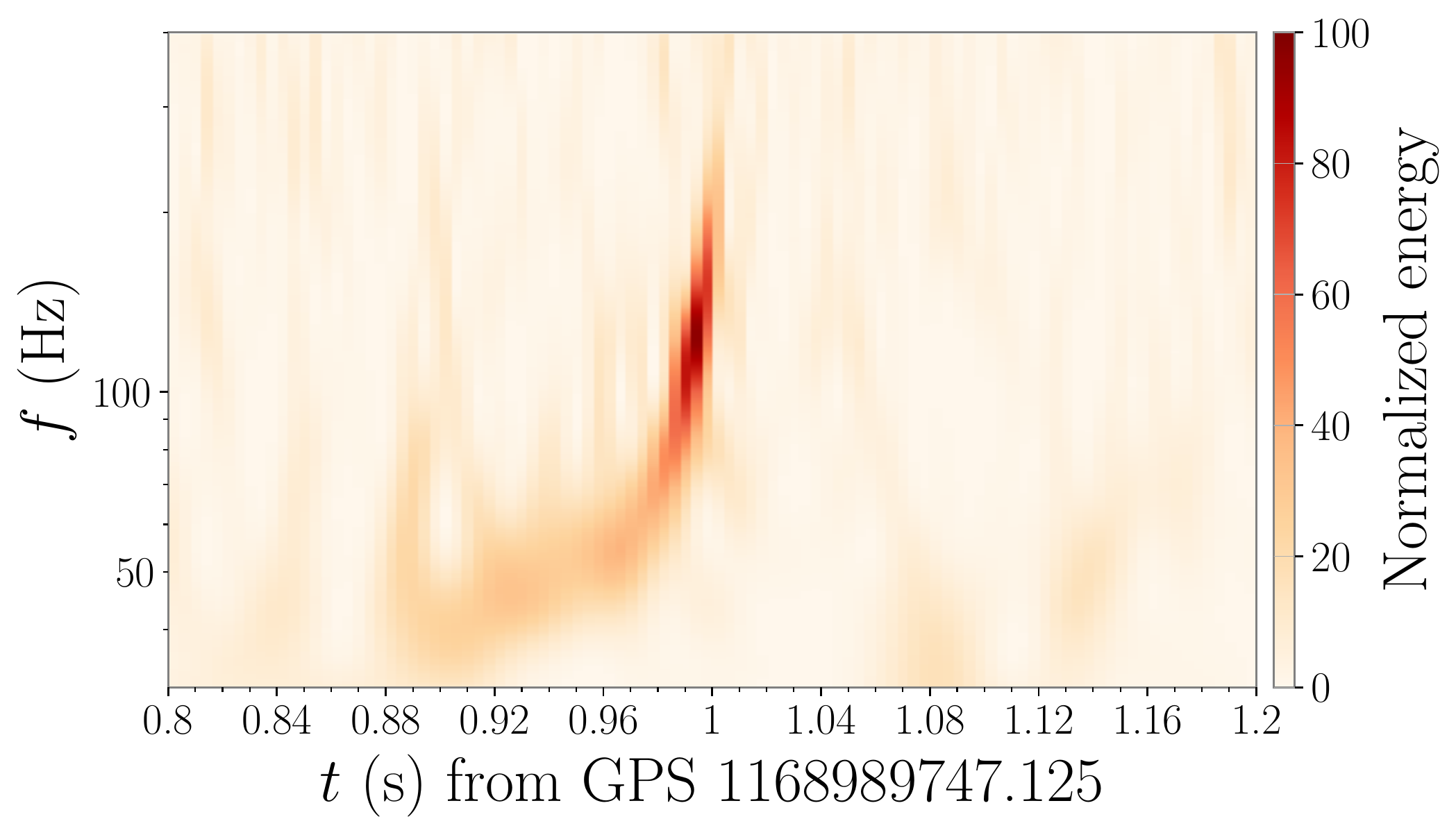}
\includegraphics[width=0.32\textwidth]{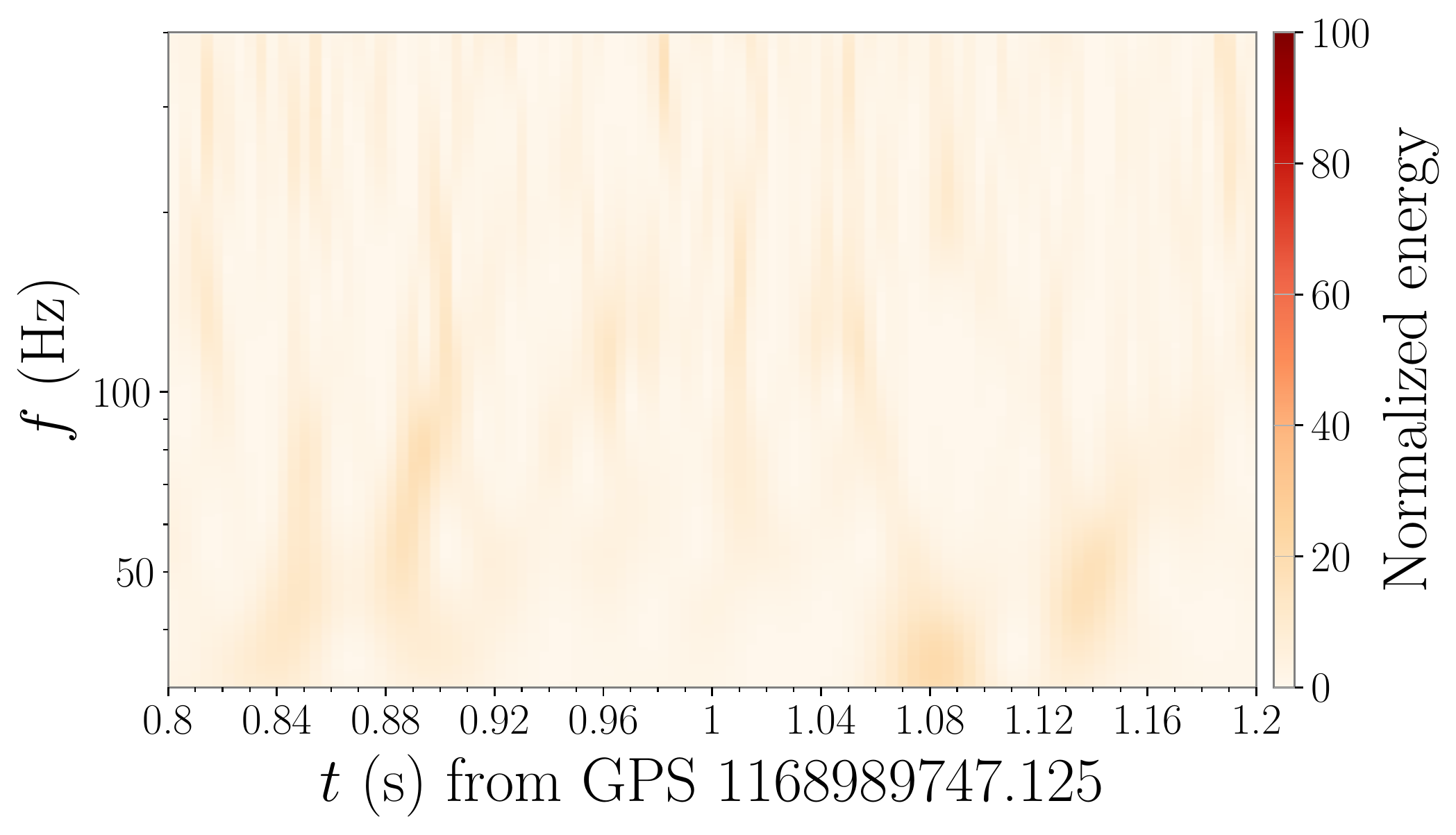}
\caption{Spectrogram of the LIGO Hanford data around the time of the blip glitch for the leftmost injection from Fig.~\ref{fig:BlipRec}.
 Left panel: data containing the blip glitch and the simulated CBC signal. Middle panel:
data after a fair draw from the glitch model has been subtracted leaving behind only the chirping CBC signal. 
Right panel: data after a fair draw from the glitch and CBC models has been subtracted, leaving behind only Gaussian detector noise.}
\label{fig:BlipSpec}
\end{figure*}

Blip glitches are one of the most common glitch types for the two LIGO detectors. They are characterized by short duration, and hence pose a challenge for the detection of high mass BBH signals~\cite{Cabero:2019orq}. Their origin 
is largely unknown. Figures~\ref{fig:BlipRec}-\ref{fig:BlipSpec} show our results for simulated signals injected at different times with respect to a blip glitch
in the LIGO Hanford detector during O2. Details about the glitch, including its GPS time, and the run settings are presented in Table~\ref{tab:settings}. A spectrogram
of the data containing the glitch is given on the left panel of Fig.~\ref{fig:SpecAll}, where the short duration and large frequency extent are shown.

The whitened data and reconstructions for the CBC signal and the glitch are shown in Fig.~\ref{fig:BlipRec} where we plot the 90\% credible intervals for each reconstruction in LIGO Hanford
(top) and LIGO Livingston (bottom). The glitch is easily visible in LIGO Hanford as a short duration $\sim 15 \sigma$ noise excursion. No glitch power is identified in 
LIGO Livingston at that time, but the CBC signal is clearly identified. This allows us to separate the corresponding coherent CBC signal in LIGO Hanford from the instrumental 
glitch, even when the latter overlaps with the merger phase of the signal (left panel). 
The glitch reconstruction is also consistent across the three simulated signals, suggesting that the glitch model is not fitting any part of the CBC signal.

Source parameters for the simulated CBC are presented in Figs.~\ref{fig:BlipParams} and~\ref{fig:BlipParams_Di} both for the CBC+glitch+noise analysis and 
a CBC+noise analysis for selected recovered parameters for the leftmost
simulated CBC signal
together with the injected values with black crosses or vertical lines as appropriate. 
Figure~\ref{fig:BlipParams} shows the mass ratio $q$, the effective spin $\chieff$, and the detector frame chirp mass ${\cal{M}}$, while Fig.~\ref{fig:BlipParams_Di}
shows the luminosity distance and the cosine of the inclination angle. 
In all cases the posterior distributions recovered under the CBC+glitch+noise model are 
consistent with the injected parameters, though the marginalized posteriors do not peak at the injected values, as expected from inference of signals in Gaussian noise. For reference, we show posteriors under the CBC+noise model in orange that assumes that the data are consisted of just a CBC signal and Gaussian noise,
without any provision for a glitch. Since this assumption is violated by the presence of the blip glitch, the resulting posteriors are expected to be biased compared to 
the true parameters and the orange contours in Figs.~\ref{fig:BlipParams} and~\ref{fig:BlipParams_Di} quantify this bias. We find that the extrinsic parameters
that are primarily determined by the signal amplitude are more biased than the intrinsic ones that are measured through the GW phase, as also discussed in~\cite{Powell:2018csz}.

The separation of the CBC signal from the glitch demonstrated in Fig.~\ref{fig:BlipRec} can be used to produce ready-to-use deglitched data for downstream inference 
analyses, as was done in~\cite{Abbott:2020niy}. An estimate of the glitch reconstruction (the median or a fair draw from the glitch model posterior) 
is subtracted from the data to produce strain data that contain only the CBC
signal and Gaussian noise. The result of the glitch subtraction is shown in the spectrograms of Fig.~\ref{fig:BlipSpec} that show the LIGO Hanford data before (left) and after 
(middle) the subtraction of a fair draw glitch reconstruction for the leftmost injection of Fig.~\ref{fig:BlipRec}. 
The left panel includes both the chirping signal and the blip glitch, while only the former is visible in the
middle panel. The right panel shows the data after a fair draw from both the CBC and the glitch models has been subtracted, resulting in residual Gaussian noise only.

\subsection{Glitch type 2: Scattered light}

Glitches caused by scattered light in the interferometer became particularly prominent during O3~\cite{Abbott:2020niy}. Unlike the blip glitches studied above, scattered light glitches have a longer
temporal duration of a few seconds and are characterized by arches in a time-frequency spectrogram~\cite{Accadia_2010,2020arXiv200714876S}, as depicted in the middle panel of Fig.~\ref{fig:SpecAll}. 
We inject simulated signals on an 
instance of such a glitch in LIGO Livingston and analyze the data from both LIGO detectors 
with our CBC+glitch+noise model. Details of the glitch and the run settings are given in Table~\ref{tab:settings}.
Due to the duration of the glitch and its low frequency power we extend our analysis duration and bandwidth. The longer duration helps the noise model determine the
low-frequency Gaussian noise PSD and thus separate the low frequency part of the glitch from Gaussian noise. We also increase the maximum quality factor $Q_{\textrm{max}}$ of 
the wavelets due to the glitch's long duration.

\begin{figure*}
\includegraphics[width=\textwidth]{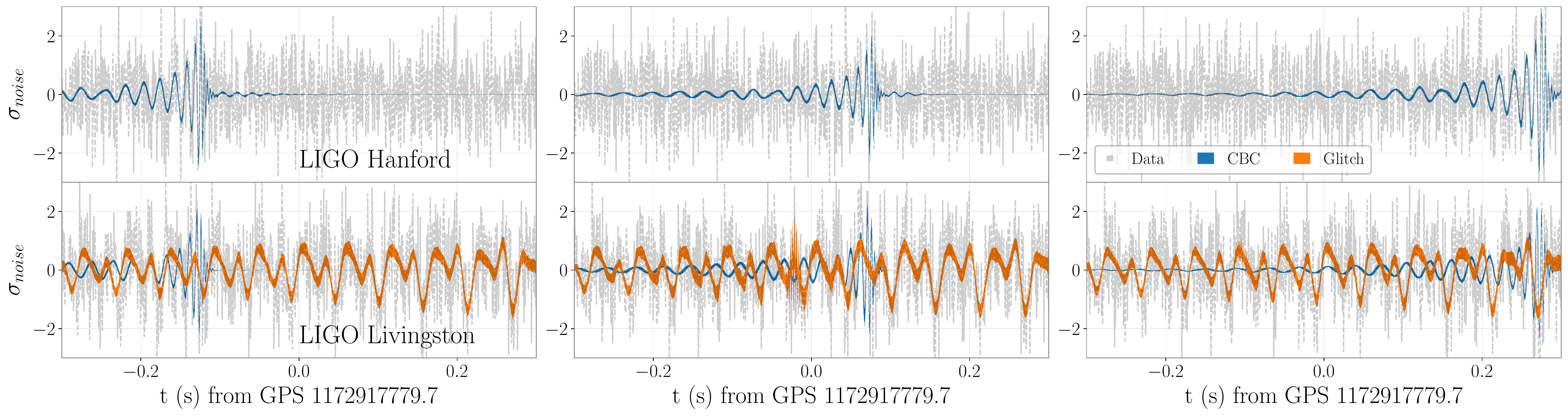}
\caption{Credible intervals for the glitch (orange) and the CBC (blue) signal reconstruction for data containing a scattered light glitch in LIGO Livingston and a simulated CBC signal at $3$ different 
times with respect to the glitch (left to right). Shaded regions correspond to 90\% credible intervals, while in grey dashed lines we plot the data whitened with a fair draw PSD from our
noise model posterior. The top row corresponds to LIGO Hanford while the bottom row corresponds to LIGO Livingston.}
\label{fig:SCRec}
\end{figure*}

\begin{figure}
\includegraphics[width=0.49\textwidth]{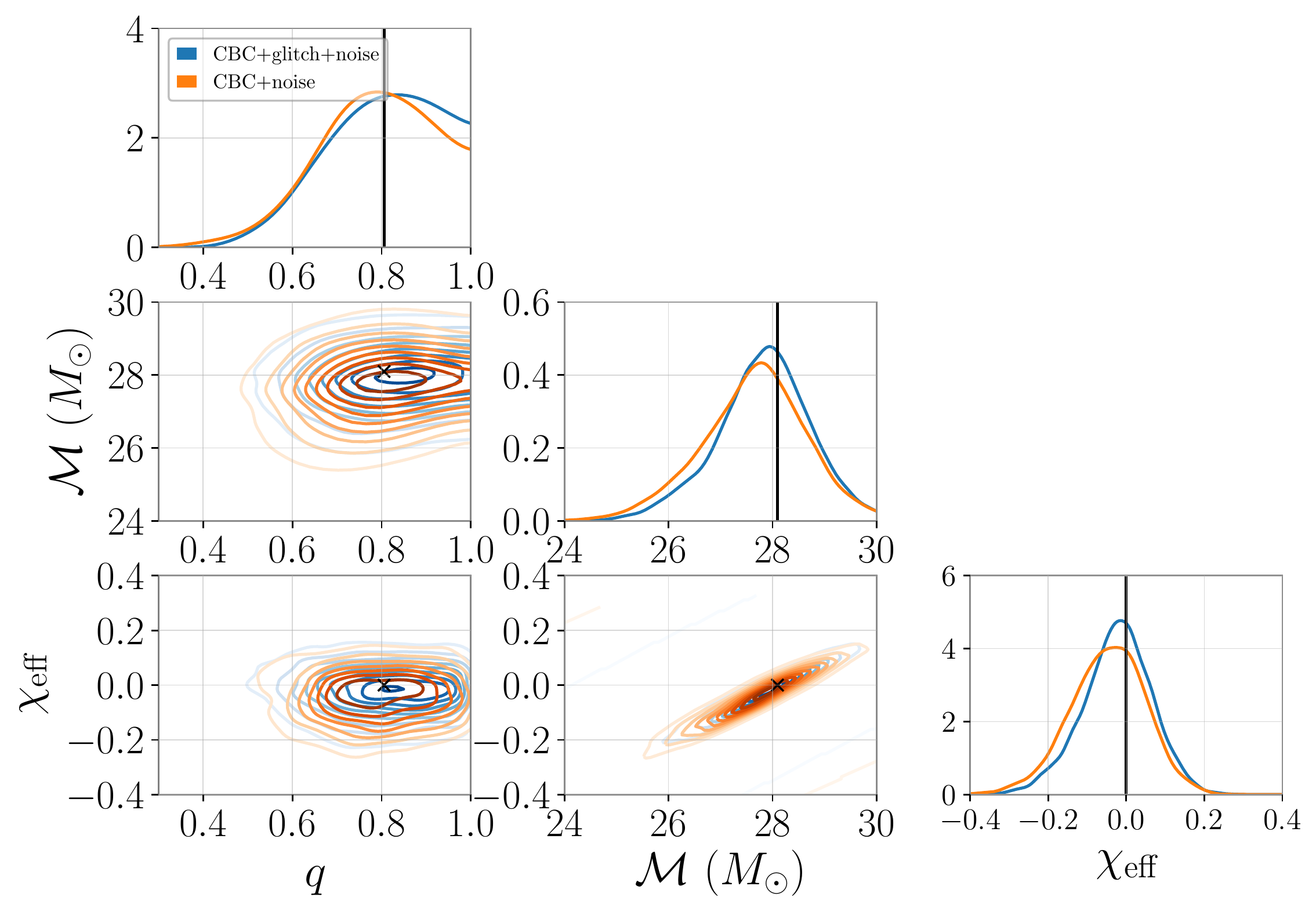}
\caption{One- and two-dimensional posterior distribution for selected source parameters of the simulated signal from the left panels of Fig.~\ref{fig:SCRec} injected on top of a
LIGO Livingston scattered light glitch. 
We include the mass ratio $q$, the effective spin $\chieff$, and the detector frame chirp mass ${\cal{M}}$ posteriors, while black crosses or black vertical lines
denote the true parameters of the injection. Blue (orange) contours and lines correspond to the CBC+glitch+noise (CBC+noise) run.}
\label{fig:SCParams}
\end{figure}

Figure~\ref{fig:SCRec} shows the data and reconstructed CBC and glitch models. We zoom in around the CBC signals, though the glitch extends beyond the time
range plotted. In all cases the CBC signal is separated from the glitch, aided by the presence of a coherent signal in LIGO Hanford. The glitch reconstruction is also consistent
for all $3$ simulated signals, as expected for runs on the same glitch. The reconstruction exhibits oscillations at around $32$Hz and $16$Hz, consistent with expectations
from the glitch spectrogram. Figure~\ref{fig:SCParams} shows posterior distributions for selected source parameters for the left-most injection in blue, 
as well as the injected 
parameters. In all cases the recovered parameters are consistent with their injected values.
In orange, we plot results from a CBC+noise run and find small biases in the source intrinsic parameters, most notably the mass ratio.

\begin{figure*}
\includegraphics[width=0.32\textwidth]{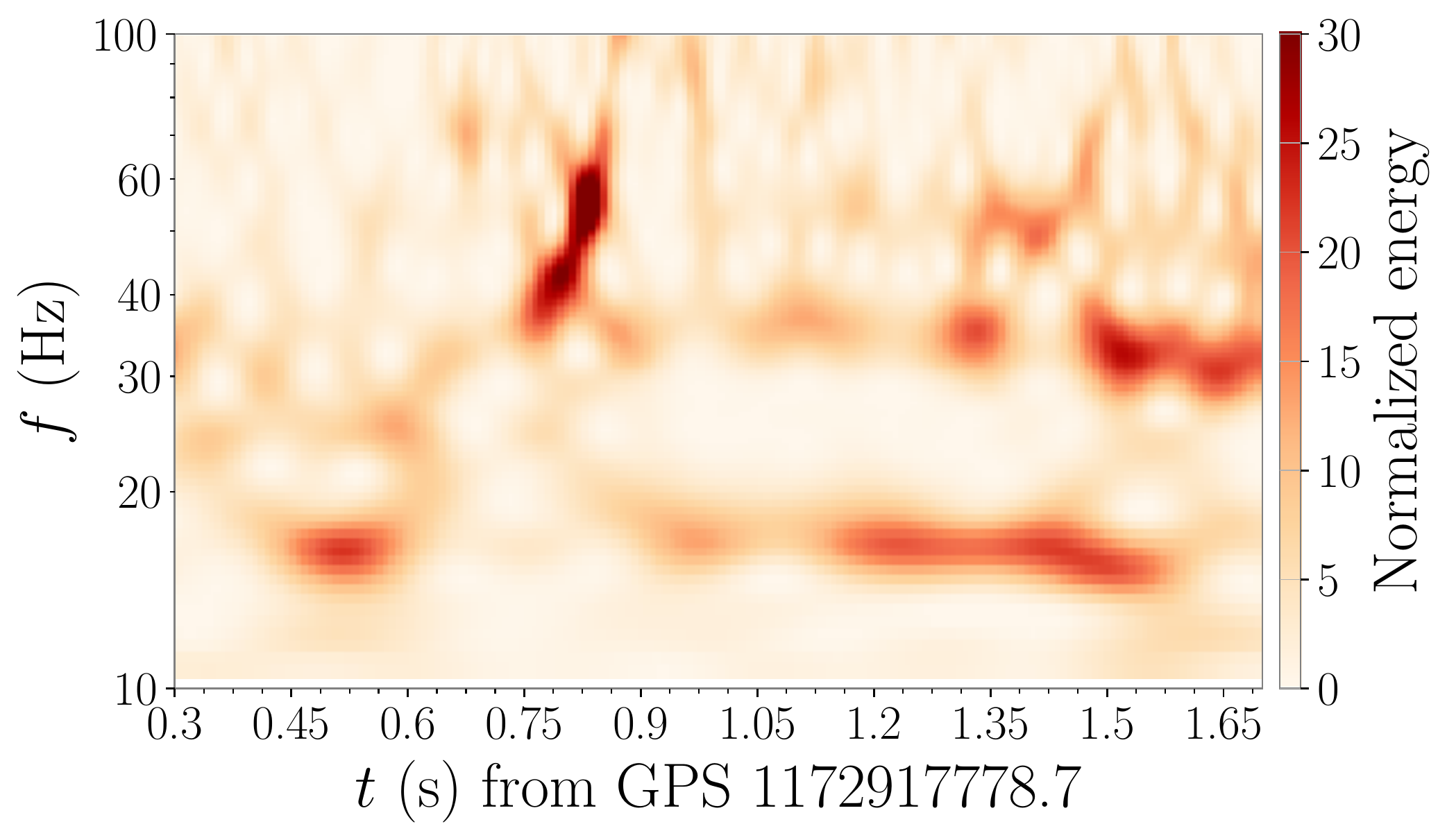}
\includegraphics[width=0.32\textwidth]{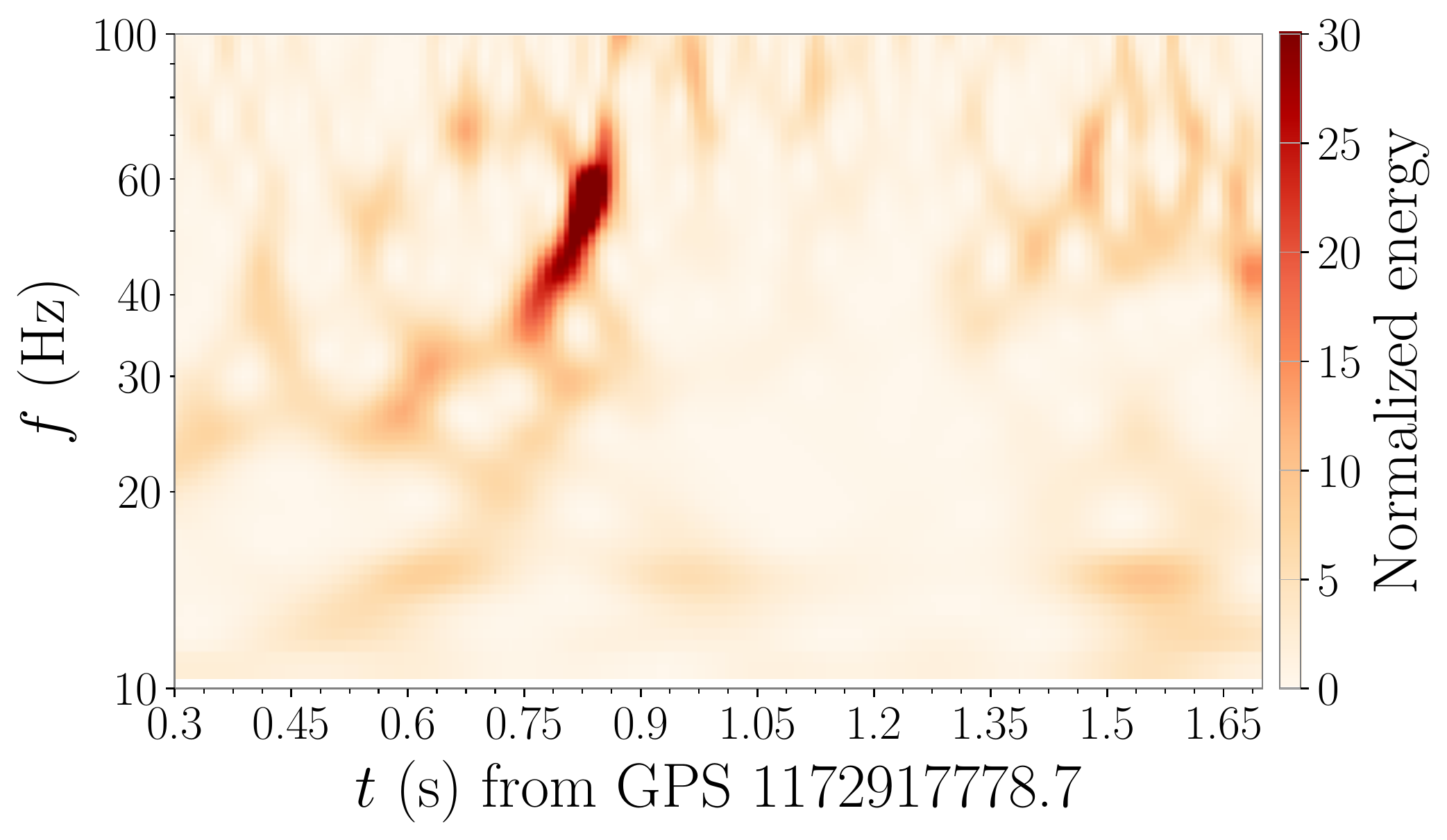}
\includegraphics[width=0.32\textwidth]{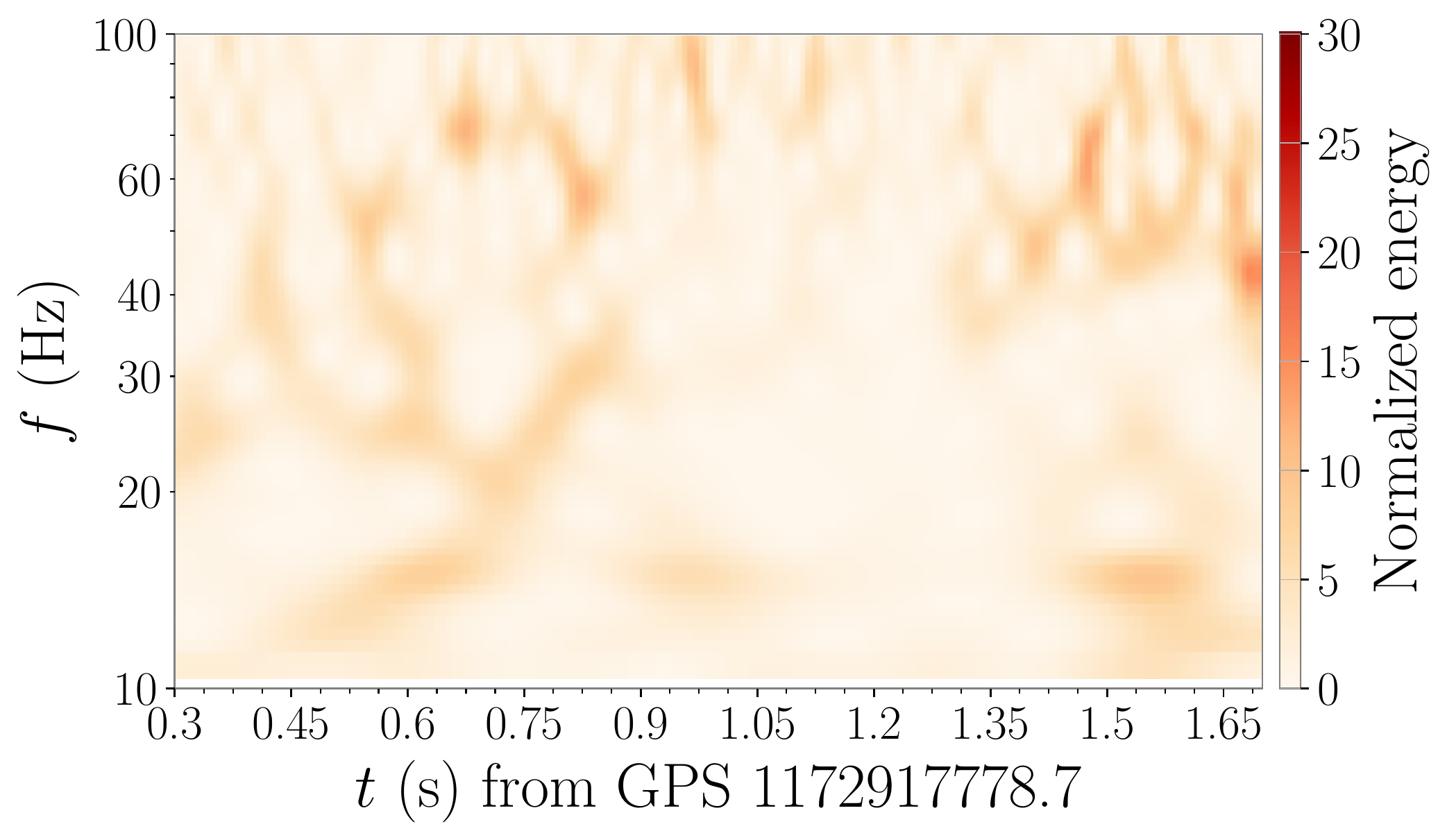}
\caption{Spectrogram of the LIGO Livingston data around the time of the scattered light glitch for the leftmost injection from Fig.~\ref{fig:SCRec}.
 Left panel: data containing the scattered light glitch and the simulated CBC signal. Middle panel:
data after a fair draw from the glitch model has been subtracted leaving behind only the chirping CBC signal. 
Right panel: data after a fair draw from the glitch and CBC models has been subtracted, leaving behind only Gaussian detector noise.}
\label{fig:SCSpec}
\end{figure*}

Finally, Fig.~\ref{fig:SCSpec} shows the spectrogram of the data before and after various components of the model have been subtracted from the data. 
The left panel corresponds to data that contain both a signal and the glitch and thus both
the signal chirp and the characteristic glitch arches are visible. In the middle panel we plot data after a fair draw from the glitch model has been subtracted,
resulting in both the high and the low frequency arches of the glitch having been regressed,
leaving only the chirping signal behind. The right panel corresponds to data where a fair draw from the CBC model has further been subtracted and is consistent
with Gaussian noise.

\subsection{Glitch type 3: Blue mountain}

\begin{figure*}
\includegraphics[width=\textwidth]{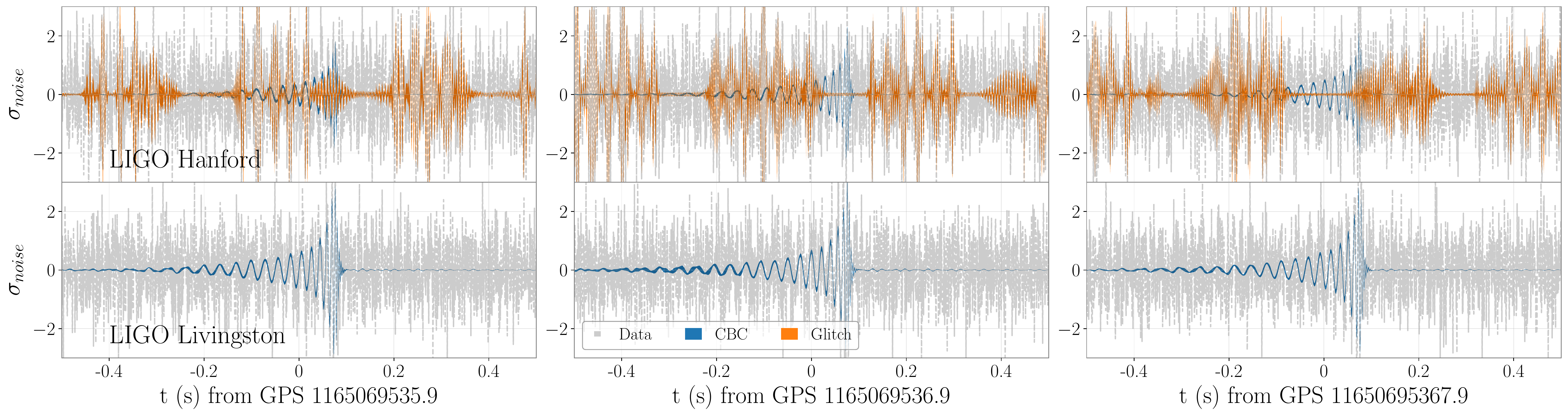}
\caption{Credible intervals for the glitch (orange) and the CBC (blue) signal reconstruction for data containing a blue mountain glitch in LIGO Hanford and a simulated CBC signal at $3$ different 
times with respect to the glitch (left to right). Shaded regions correspond to 90\% credible intervals, while in grey dashed lines we plot the data whitened with a fair draw PSD from our
noise model posterior. The top row corresponds to LIGO Hanford while the bottom row corresponds to LIGO Livingston.}
\label{fig:BMRec}
\end{figure*}

\begin{figure}
\includegraphics[width=0.49\textwidth]{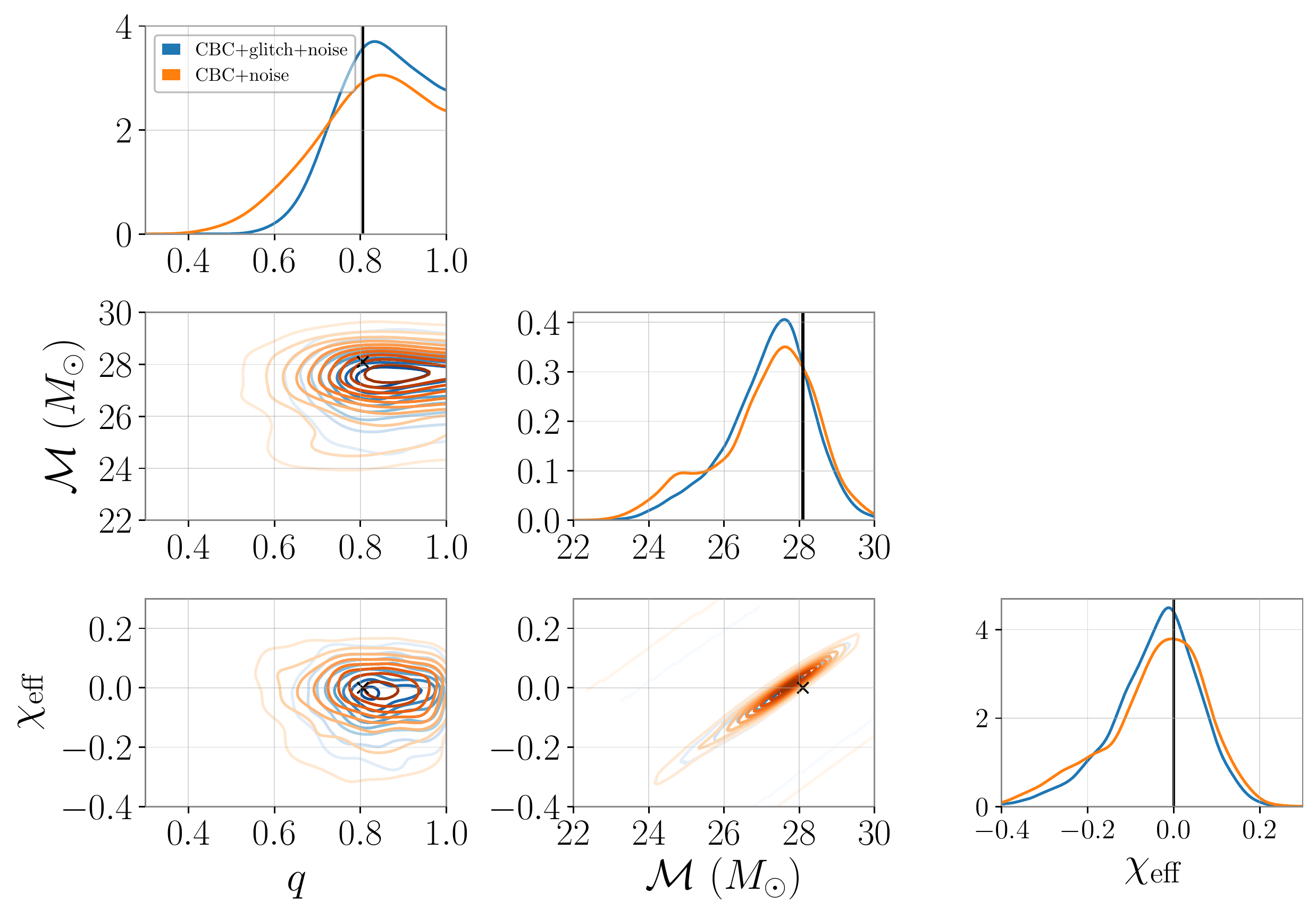}
\caption{One- and two-dimensional posterior distribution for selected source parameters of the simulated signal from the left panels of Fig.~\ref{fig:BMRec} injected on top of a
LIGO Hanford blue mountain glitch. 
We include the mass ratio $q$, the effective spin $\chieff$, and the detector frame chirp mass ${\cal{M}}$ posteriors, while black crosses or black vertical lines
denote the true parameters of the injection.}
\label{fig:BMParams}
\end{figure}

The final type of glitch we consider is the blue mountain; the spectrogram of the LIGO Hanford instance of a blue mountain glitch we consider is shown 
 in the right panel of Fig.~\ref{fig:SpecAll}. The glitch has a duration of multiple
seconds and is characterized by higher frequencies $\sim 200$Hz. We inject simulated signals at different times relative to the glitch and again analyze data from the two LIGO
detectors with
the CBC+glitch+noise model with settings shown in Table~\ref{tab:settings}. Due to the large glitch duration we have to increase the length of the analyzed segment
even further to $16$s.
Despite the glitch's overall long duration, we do not find it necessary to increase the wavelet maximum quality factor $Q_{\textrm{max}}$, as the glitch is composed of short
individual bursts of power, each of which is modeled by individual wavelets with a small quality factor.

Figure~\ref{fig:BMRec} shows the whitened data and credible intervals for the whitened CBC and glitch reconstruction in each detector for each of the injected
signals. Due to the large glitch duration, the signals are injected sufficiently wide apart that the reconstruction plots show non overlapping parts of the data
and the glitch. The glitch reconstructions are therefore not expected to match. As expected from the glitch spectrogram, the glitch is characterized by a series of
short high frequency bursts, each of which is modeled by different wavelets within our glitch model. 
Figure~\ref{fig:BMParams} shows posterior distributions for selected source parameters for the left-most injection in blue, as well as the injected 
parameters. In all cases the 
recovered parameters are consistent with their injected values, suggesting that the presence of the glitch does not 
incur biases on the inferred source properties if the two are modeled simultaneously.
As before, we also plot results from a CBC+noise run that neglects the glitch in the data in orange and again find small biases by the presence of the glitch 
in the source intrinsic parameters.

\begin{figure*}
\includegraphics[width=0.32\textwidth]{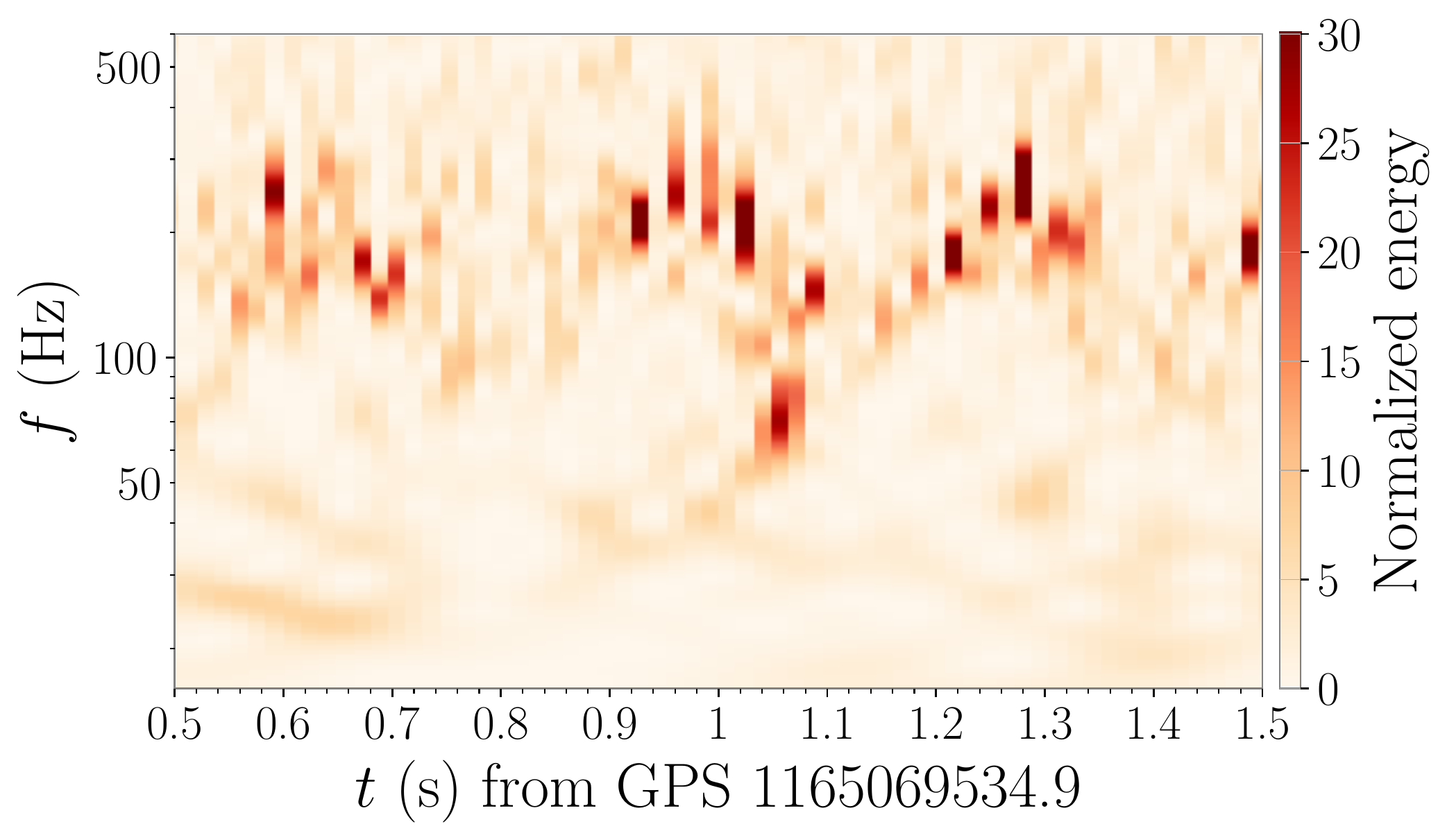}
\includegraphics[width=0.32\textwidth]{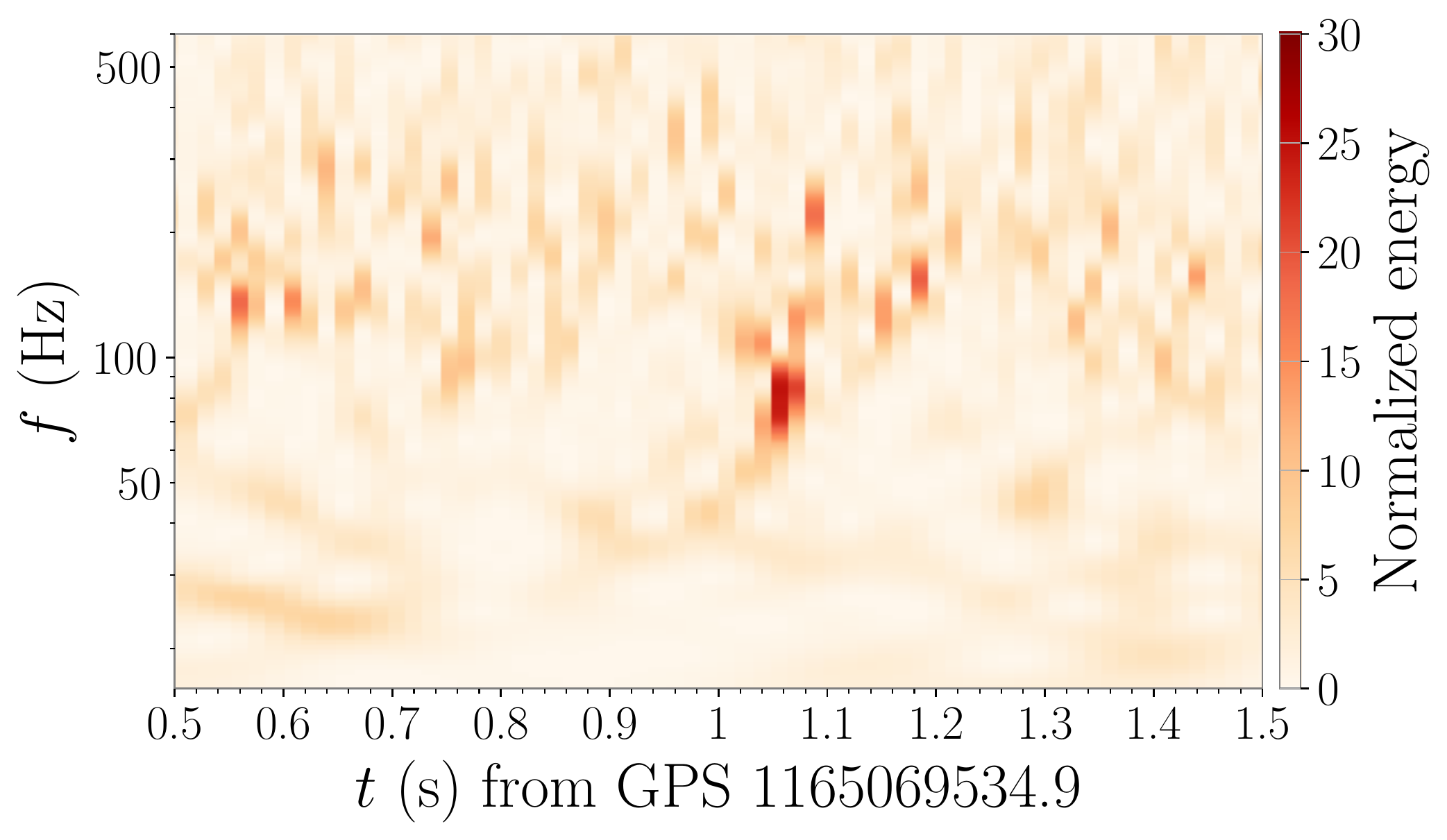}
\includegraphics[width=0.32\textwidth]{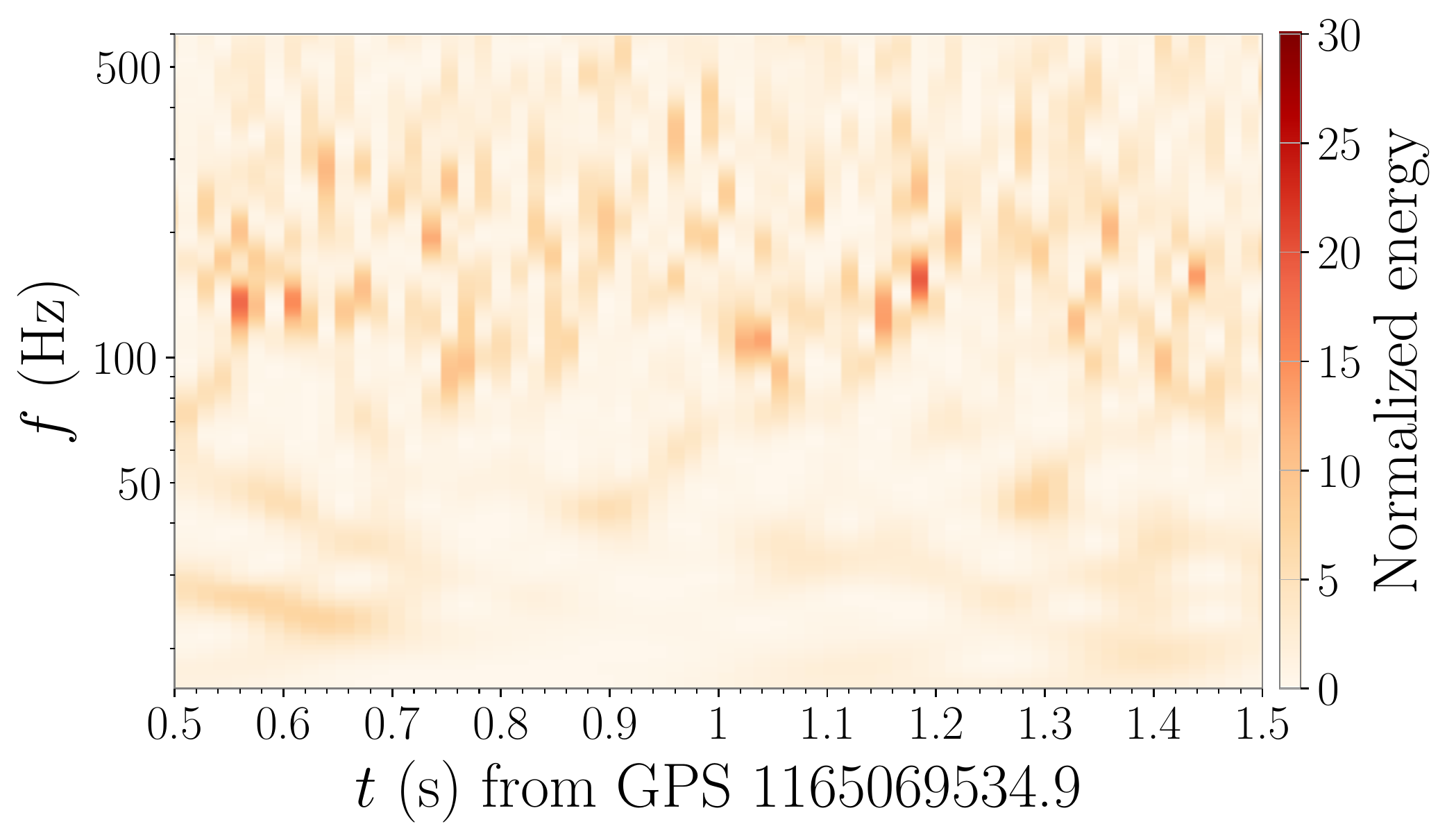}
\caption{Spectrogram of the LIGO Livingston data around the time of the blue mountain glitch for the leftmost injection from Fig.~\ref{fig:BMRec}.
 Left panel: data containing the blue mountain glitch and the simulated CBC signal. Middle panel:
data after a fair draw from the glitch model has been subtracted leaving behind only the chirping CBC signal. 
Right panel: data after a fair draw from the glitch and CBC models has been subtracted, leaving behind only Gaussian detector noise.}
\label{fig:BMSpec}
\end{figure*}

The glitch subtraction process is detailed in Fig.~\ref{fig:BMSpec} that again shows spectrograms of the original data containing both the glitch and the signal (left),
data after a fair draw glitch model has been subtracted (middle), and data after both the glitch and the fair draw CBC model have been removed (right). As before,
data from the middle panel could be used for further data processing. The right panel shows data where a model for both the glitch and the CBC have been subtracted. 
Even though the majority of the glitch power is absent (compare the left and right panels), some small non Gaussian power might be left behind. The reason for this is that
the blue mountain glitch is manifested as individual short bursts of glitch power, which our flexible analysis attempts to model completely independently. Indeed, the glitch model
for this run uses ${\cal{O}}(70)$ wavelets. Each of these wavelets, needs to model sufficient non-Gaussian power in the data in order to overcome the parsimony penalty incurred by
adding more parameters to the model. As such, we expect that some of the weaker ``bursts" of the glitch will not be recovered. Possible ways to alleviate this are discussed in Sec.~\ref{sec:conclusions}.

\section{Gravitational Wave Events}
\label{sec:events}

As a further demonstration of our CBC+glitch+noise model, we also analyze two astrophysical events, GW170817~\cite{TheLIGOScientific:2017qsa} 
and GW150914~\cite{Abbott:2016blz} whose data are available from GWOSC~\cite{GWOSC,Abbott:2019ebz}. 
Though not the main focus of this paper, the analysis presented below also provides an estimate of the effect of marginalizing
over the noise PSD has on the inferred astrophysical parameters. More details about this effect will be presented in a separate study.

\subsection{GW170817}

Perhaps the most known instance of a GW signal overlapping with an instrumental glitch is GW170817~\cite{TheLIGOScientific:2017qsa}. 
Inference on the GW170817 source properties is performed on data where the glitch in LIGO Livingston has been modeled with \BayesWave's glitch-only model and subtracted. Analysis of simulated signals suggests that
this procedure leads to unbiased inference, while any analysis on data that contain the glitch results in highly biased source parameters~\cite{Pankow:2018qpo}. Both versions of the data are publicly available, both with and 
without the glitch~\cite{GW170817Data}, so we analyze them both with different models. We use data from the LIGO Hanford and the LIGO Livingston detectors and analyze $64$s of data from $16$Hz to $2048$Hz using the {\tt IMRPhenomD\_NRTides} waveform model that includes finite-size effects~\cite{Dietrich:2018uni}. We employ our
CBC+glitch+noise model on the data with the glitch and the CBC+noise model on data where the glitch has already been subtracted. For the CBC+glitch+noise case we use \GlitchBuster~\cite{Cornish:2020dwh} to provide a quick fit
to the glitch and use that as a starting point for our glitch model during sampling.

\begin{figure}
\includegraphics[width=0.49\textwidth]{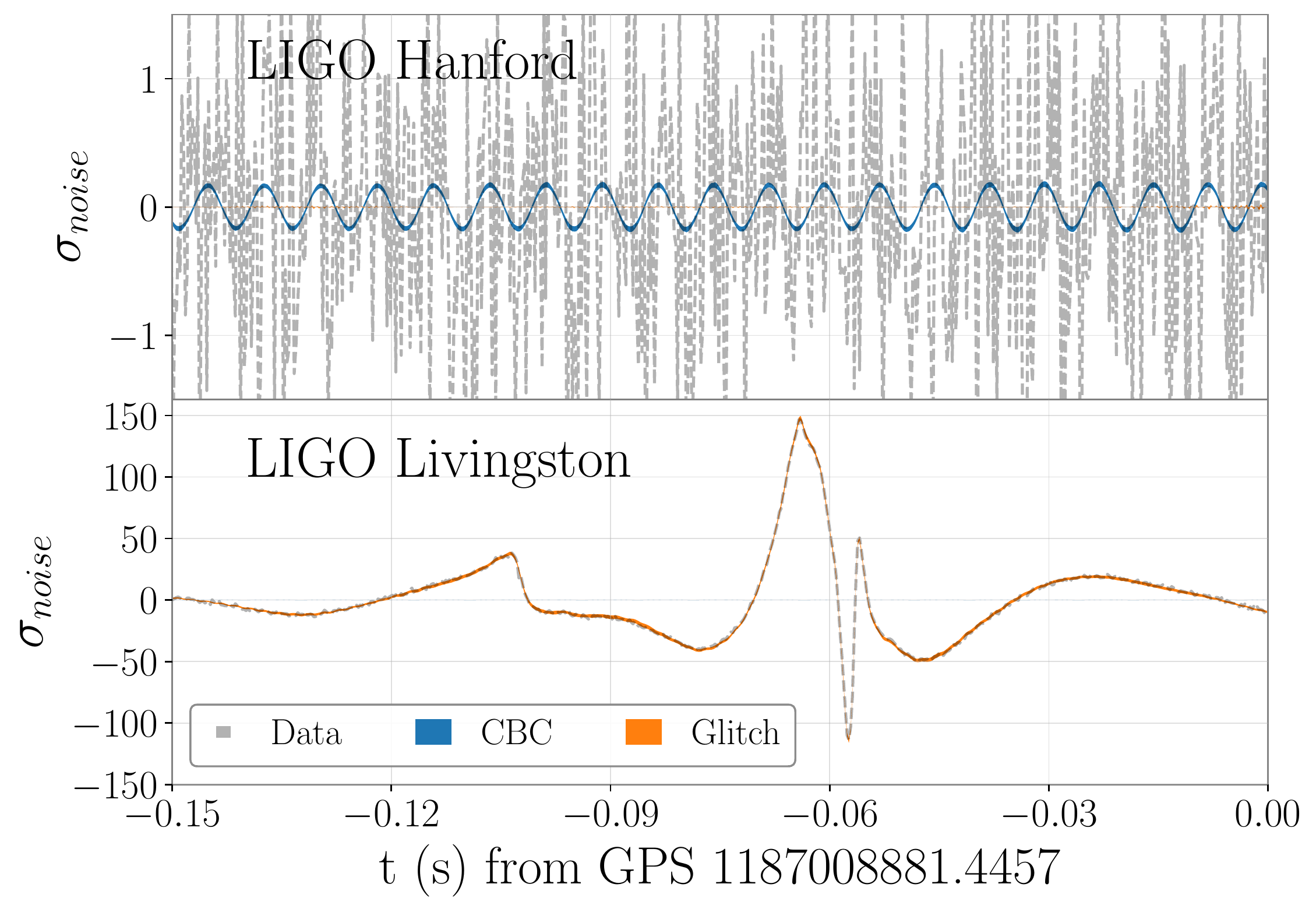}
\caption{Credible intervals for the glitch (orange) and the CBC (blue) signal reconstruction for GW170817. 
Shaded regions correspond to 90\% credible intervals, while in grey dashed lines we plot the data whitened with a fair draw PSD from our
noise model posterior. The top row corresponds to LIGO Hanford while the bottom row corresponds to LIGO Livingston. The LIGO Hanford plot zooms in to show the signal that is invisible in the LIGO Livingston plot due to
the size of the glitch; note the y-scale difference in the two plots}
\label{fig:170817rec}
\end{figure}

\begin{figure}
\includegraphics[width=0.49\textwidth]{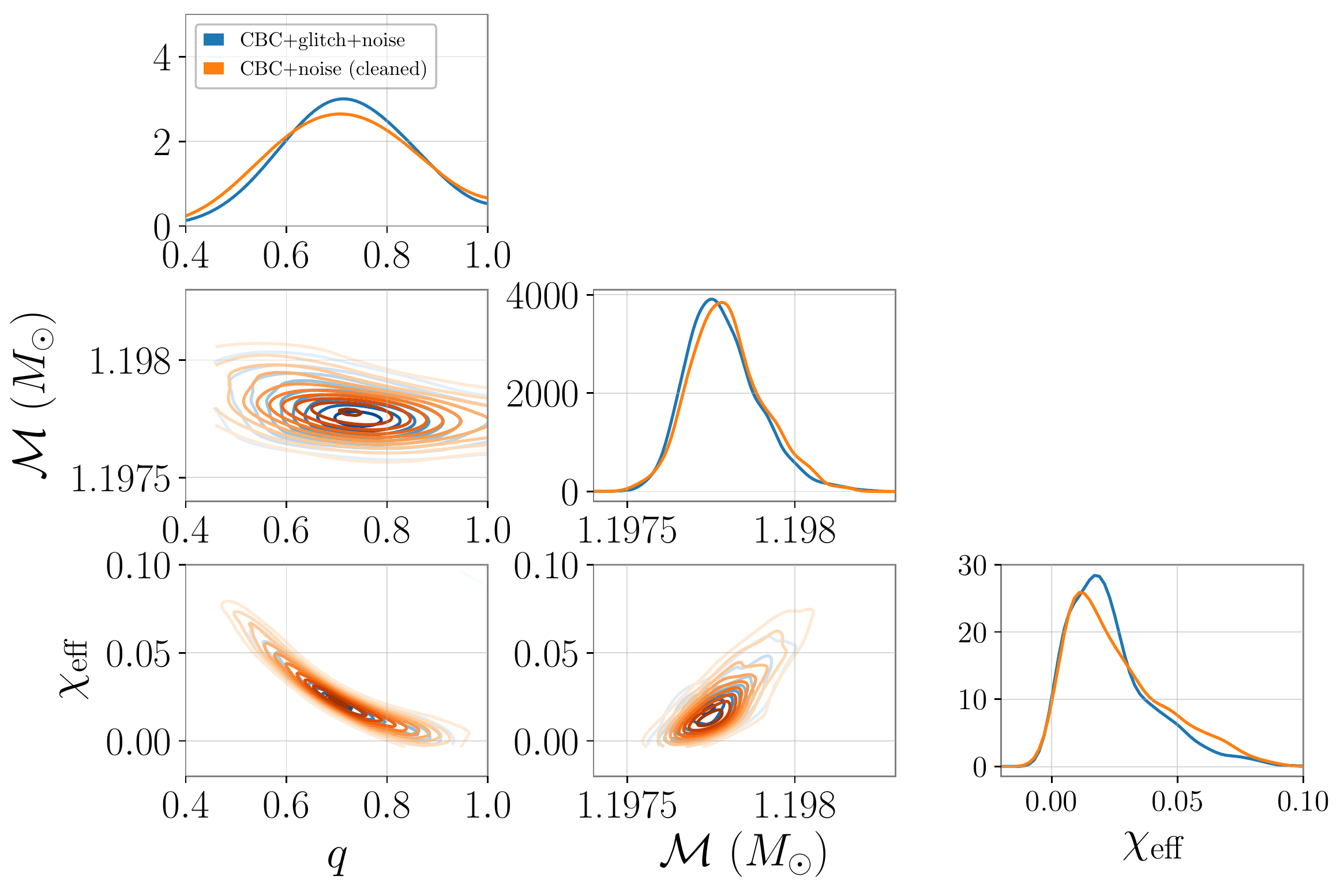}
\caption{One- and two-dimensional posterior distribution for selected source parameters for GW170817. 
We include the mass ratio $q$, the effective spin $\chieff$, and the detector frame chirp mass ${\cal{M}}$ posteriors. Blue curves show posteriors under the
CBC+glitch+noise model on the full data, while orange curves correspond to the CBC+noise model on data where the glitch has already been subtracted. The two sets of results are consistent with each other.}
\label{fig:170817pe}
\end{figure}

Credible intervals for the signal and glitch reconstructions are shown in Fig.~\ref{fig:170817rec} for each detector for $\sim150$ms of data around the glitch. Despite its high SNR, GW170817 had a relatively low amplitude, so the LIGO Hanford
plot has been zoomed in to make the signal visible. The LIGO Livingston data  are dominated by the glitch, peaking at $\sim 150\sigma$ relative to the background detector Gaussian noise. The signal is not visible in the
LIGO Livingston data given the plotting scale. Figure~\ref{fig:170817pe} shows selected source parameters obtained from data both with and without the glitch. We find consistent results, showing that our combined CBC+glitch+noise analysis can 
faithfully fit the CBC signal and the glitch simultaneously, without the need for the two step process of first removing the glitch and then reanalyzing the data.

\subsection{GW150914}

The first GW signal directly detected by the LIGO detectors, GW150914~\cite{Abbott:2016blz}, did not overlap with an instrumental glitch~\cite{TheLIGOScientific:2016zmo}. 
However, since it is one of the best studied
and loudest signals, we select it as a demonstration of our analysis on data without glitches. Our glitch model has the flexibility to use no glitch wavelets, we therefore expect
many samples in the glitch model posterior to contain exactly zero glitch power. We analyze $4$s of data starting at $16$Hz and with a sampling rate of $2048$Hz. We perform two
runs, one with the CBC+glitch+noise model and one with the CBC+glitch model using otherwise identical settings.

\begin{figure}
\includegraphics[width=0.49\textwidth]{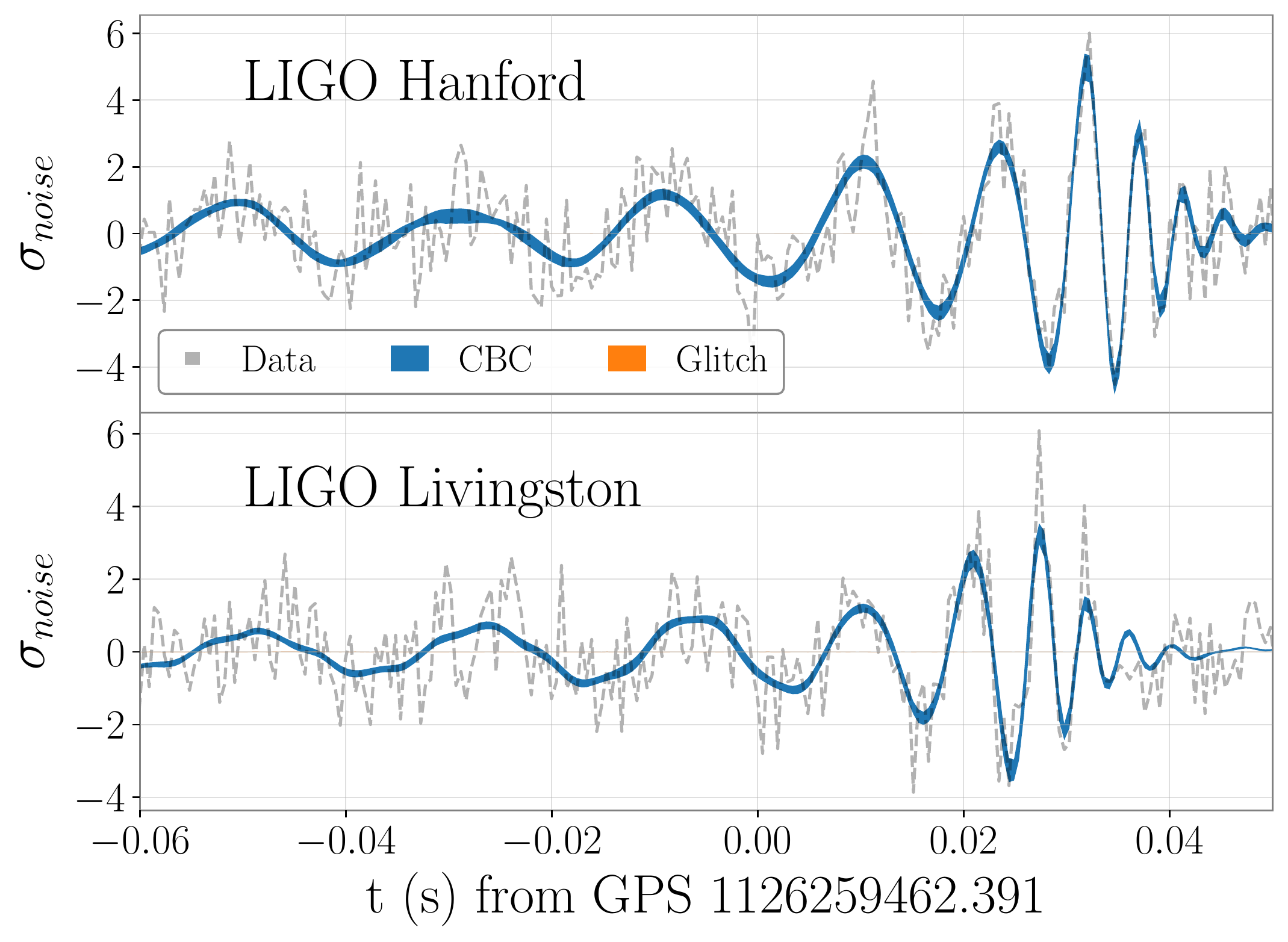}
\caption{Credible intervals for the glitch (orange) and the CBC (blue) signal reconstruction for GW150914. 
Shaded regions correspond to 90\% credible intervals, while in grey dashed lines we plot the data whitened with a fair draw PSD from our
noise model posterior. The top row corresponds to LIGO Hanford while the bottom row corresponds to LIGO Livingston. Our glitch model recovers essentially no incoherent power
coincident with the astrophysical signal and therefore the reconstruction is not visible in the scale of the plot.}
\label{fig:150914rec}
\end{figure}

\begin{figure}
\includegraphics[width=0.49\textwidth]{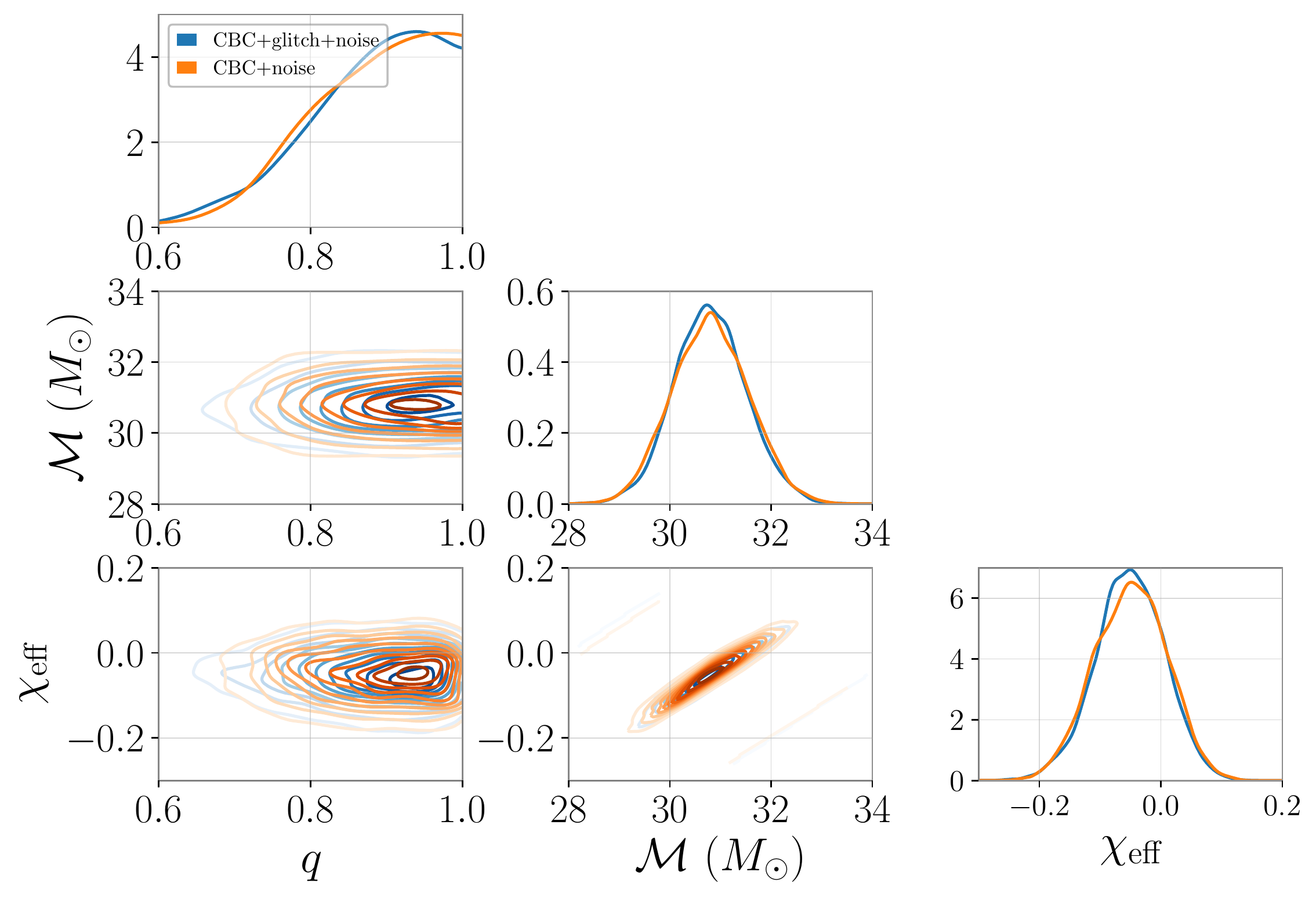}
\caption{One- and two-dimensional posterior distribution for selected source parameters for GW150914. 
We include the mass ratio $q$, the effective spin $\chieff$, and the detector frame chirp mass ${\cal{M}}$ posteriors. Blue curves show posteriors under the
CBC+glitch+noise model, while orange curves correspond to the CBC+noise model. The two sets of results are consistent with each other.}
\label{fig:150914pe}
\end{figure}

Relevant results are shown in Figs.~\ref{fig:150914rec} and~\ref{fig:150914pe} where as before
we plot the CBC and glitch reconstructions of the CBC+glitch+noise model in the two detectors and the recovered source parameters. 
The CBC reconstruction of Fig.~\ref{fig:150914rec} is consistent with 
previous results~\cite{TheLIGOScientific:2016wfe}. The glitch reconstruction is too small to identify in the scale of the plot, as we find that $86\%$ and $14\%$ of our posterior
samples had exactly zero glitch wavelets in LIGO Hanford and LIGO Livingston respectively. Figure~\ref{fig:150914pe} shows the posterior distribution
 for selected source parameters of 
GW150914 obtained under the CBC+glitch+noise and the CBC+noise models. The two posteriors yield consistent results, 
showing that the glitch model does not affect the CBC parameters
when no glitch is present in the data.

\section{Conclusions}
\label{sec:conclusions}

We construct and validate an analysis of GW data that simultaneously models astrophysical CBC signals and instrumental glitches. We test the analysis against real instances of glitches in the two LIGO detectors
from O2 data and simulated CBC signals injected at different times with respect to the glitch. We find that our analysis can separate the two, and provide both estimates for the CBC source parameters and 
glitch-subtracted data for subsequent analyses. The glitch model we employ is a sum of sine-Gaussian wavelets that is not tuned to any specific glitch type and morphology; it can thus handle even novel
glitch types that might first appear during O4. Even though this flexibility is desirable given the unpredictable and evolving nature of glitches, the efficacy of glitch subtraction can be improved by employing targeted
priors for different glitch types. One such example would be a prior that anticipates arches at frequency multiples in the case of scattered light glitches. We leave such targeted priors to future work.

Our analysis considered only simulated BBH signals, though we also present an analysis of the BNS GW170817. We expect overlapping CBCs and glitches of similar duration to be a
worse-case-scenario due to their similar morphology~\cite{TheLIGOScientific:2017lwt,Davis:2020nyf}. Given that, we plan to carry out a larger scale study of our CBC+glitch analysis that includes more glitch types and CBC classes, 
such as BNSs and lower mass BBHs. Additionally, the analysis presented here did not make use of \GlitchBuster~\cite{Cornish:2020dwh} to provide initial fits to the glitch, apart from the GW170817 case.
In the future we plan to investigate interfacing \GlitchBuster and \BayesWave in more detail, in the hopes that an efficient starting point for the glitch model during sampling will decrease the sampler's convergence time
and result in ready-to-use glitch-subtracted data more quickly.
We hope that our analysis will contribute to robust and efficient glitch mitigation against the increased event rate anticipated in O4; our goal is to facilitate analysis of as much data as possible and
maximize the science output of the upcoming observations.

\acknowledgements

We thank Derek Davis, Laura Nuttall, and Jessica McIver for sharing preliminary results and datasets for LIGO glitches and CBC injections.
This research has made use of data, software and/or web tools obtained from the Gravitational Wave Open Science Center (https://www.gw-openscience.org), a service of LIGO Laboratory, the LIGO Scientific Collaboration and the Virgo Collaboration.
LIGO is funded by the U.S. National Science Foundation.
Virgo is funded by the French Centre National de Recherche Scientifique (CNRS), the Italian Istituto Nazionale della Fisica Nucleare (INFN) and the Dutch Nikhef, with contributions by Polish and Hungarian institutes.
The authors are grateful for computational resources provided by the LIGO Laboratory and supported by National Science Foundation Grants PHY-0757058 and PHY-0823459.
N.J.C. appreciates the support provided by NSF award PHY-1912053.
M.W. gratefully acknowledges support and hospitality from the Simons Foundation through the pre-doctoral program at the Center for Computational Astrophysics, Flatiron Institute.
The Flatiron Institute is supported by the Simons Foundation.
Software: gwpy~\cite{duncan_macleod_2020_3598469}, matplotlib~\cite{Hunter:2007}.
\bibliography{OurRefs}

\end{document}